%% file: sukhdeep_EG.tex
\documentclass[a4paper,fleqn,usenatbib]{mnras}

\usepackage{natbib}
\usepackage{multicol}

\usepackage{graphicx}	
\usepackage{amsmath}	
\usepackage{amssymb}	
\usepackage{subcaption}
\usepackage{float}

\DeclareGraphicsExtensions{.pdf,.png,.jpg,.eps}
\graphicspath{{dr12_figures/}}
\input{def.tex}

\usepackage{color}
\definecolor{sdeep}{rgb}{0., 0.6, 0.}
\definecolor{refR}{rgb}{0.5, 0.6, 0.}
\newcommand{\referee}[1]{{{#1}}}
\newcommand{\refereee}[1]{{{#1}}}

\renewcommand{\vec}[1]{\mathbf{#1}}

\title[Probing gravity with $E_G$]{
Probing gravity with a joint analysis of galaxy and CMB lensing and SDSS spectroscopy}

\author[S.~Singh et al.]{
   Sukhdeep Singh$^{1,2,3,4}$\thanks{E-mail: sukhdeep1@berkeley.edu},
   Shadab Alam$^{5,1}$, 
   Rachel Mandelbaum$^{1}$, 
   Uro\v{s} Seljak$^{2,3,4}$,\newauthor
   Sergio Rodriguez-Torres$^{6,7}$,
   Shirley Ho$^{4,3,2,1}$
\\
   $^{1}$ McWilliams
   Center for Cosmology, Department of Physics, Carnegie Mellon
   University, Pittsburgh, PA 15213, USA\\
   $^{2}$ Berkeley Center for Cosmological Physics, University of California, Berkeley, CA 94720, USA\\
   $^{3}$ Department of Physics, University of California, Berkeley, CA 94720, USA\\
   $^{4}$Lawrence Berkeley National Laboratory (LBNL),
Physics Division, Berkeley, CA 94720-8153, USA\\
   $^{5}$ Institute for Astronomy, University of Edinburgh, Royal Observatory, Blackford Hill, Edinburgh, EH9 3HJ , UK\\
   $^{6}$ Departamento de F\'isica Te\'orica M8, Universidad Aut\'onoma de Madrid (UAM), Cantoblanco, E-28049, Madrid, Spain\\
   $^{7}$ Instituto de F\'isica Te\'orica, (UAM/CSIC), Universidad Aut\'onoma de Madrid, Cantoblanco, E-28049 Madrid, Spain\\
}

\date{Accepted XXX. Received YYY; in original form ZZZ}

\pubyear{2018}

\begin{document}
\label{firstpage}
\pagerange{\pageref{firstpage}--\pageref{lastpage}}
\maketitle

\begin{abstract}
	We present measurements of \eg, a probe of gravity from large-scale structure, using BOSS LOWZ and CMASS spectroscopic 
	samples, with 
	lensing measurements from SDSS (galaxy lensing)
	and Planck (CMB lensing). Using SDSS lensing and the BOSS LOWZ sample, we measure
    \referee{$\mean{\eg}=0.40^{+0.05}_{-0.04}$ (stat), 
	 consistent with the predicted value from the Planck \lcdm\ model, $\eg=0.46$.}
	Using CMB lensing, we measure
	$\mean{\eg}=0.46^{+0.08}_{-0.09}$ (stat) for LOWZ (statistically consistent with galaxy lensing and
     Planck predictions) and 
	$\mean{\eg}=0.39^{+0.05}_{-0.05}$ (stat) for the CMASS sample, consistent with the Planck prediction of
    $\eg=0.40$ given the higher redshift of the sample.
	We also study the redshift evolution of \eg\ by splitting the LOWZ sample into two samples based on redshift, with 
	results being consistent with model predictions. 
\referee{We estimate systematic uncertainties on the above $\mean{\eg}$ numbers to be $\sim 6$\% (when using galaxy-galaxy
lensing) or $\sim 3$\% (when using CMB lensing), subdominant to the quoted statistical errors.
These systematic error budgets are dominated by observational systematics in galaxy-galaxy lensing
and by theoretical modeling uncertainties, respectively. We do not estimate observational systematics in galaxy-CMB lensing cross correlations.}
\end{abstract}


\section{Introduction}
	The standard $\Lambda$CDM model of cosmology has been successful in explaining a wide array of cosmological 
	measurements \cite[see][for a review]{Weinberg2013}, 
	from  the early Universe \citep[e.g.,][]{Steigman2010,Komatsu2011,Planck2015cosmo} down to 
	$z\lesssim1$  
	\citep[e.g.,][]{Riess1998,Perlmutter1999,Kilbinger2013,Mandelbaum2013,Betoule2014,Boss2016combined},
	 though there are some mild tensions between different probes 
	\citep[see for example][]{Planck2015cosmo,Riess2016}. 
	General relativity (GR) lies at the core of this model but it requires additional 
	matter and energy components (dark matter and dark energy) to explain structure formation and cosmic 
	acceleration. 
	The nature of these components, especially dark energy, is not very well understood and this
    leaves open the 
	possibility that the laws of gravity may require modifications as well 
	\citep[e.g.,][]{Jain2010}.

	The theory of general relativity has been 
	remarkably successful in explaining results over a wide range of scales
	\citep[see][for a review of experimental tests of GR]{Will2014}.
	On cosmological scales, it is possible to test the nature of gravity through 
	several observables since gravity determines the dynamics and the growth of structure.
	
	One important probe is the 
	large scale velocity field, particularly 
	the redshift space distortions \citep[see][for a review]{Hamilton1997}. 
	Observationally, local motions (or peculiar velocities) of galaxies introduce errors
	in the distances inferred using the cosmological redshift-distance relation. 
	Due to the coherent nature of the velocities, these errors then lead to detectable distortions
	in the otherwise isotropic correlation function (or power spectrum in Fourier space) of galaxies. These distortions 
	in the redshift-space correlation function depend on the strength of gravitational attraction and are parameterized 
	through the redshift- (or time-) dependent growth rate factor, $f(z)$. Several surveys have detected RSD and 
	constrained $f(z)$ at different redshifts \citep[e.g.,][]{Beutler2012,Torre2013,AlamRSD2015,Boss2016combined} 
	and used it to constrain gravity \citep{Jennings2011,Alam2016LCDM}.
	
	Gravitational lensing is another probe of the large scale structure. Gravitational lensing results from deflections 
	in the path of photons by the gravitational potential of intervening matter, as they travel from the source to the 
	observer \citep{Bartelmann2001,Kilbinger2015}. 
	In the weak regime, lensing introduces small but coherent distortions in the shapes of galaxies. Correlations 
	between the shapes of background galaxies can thus be used to study the gravitational potential of
    foreground matter. Similarly, in the case of the 
	CMB, lensing remaps the background anisotropies, leading to cross-correlations between different modes that can be 
	exploited to recover the matter potential \citep{Zaldarriaga1999,Hu2001,Lewis2006}. 
	Cross-correlations between galaxies and the lensing maps from CMB (galaxy-CMB lensing) 
	or background galaxies (galaxy-galaxy lensing) can be used to study the galaxy-matter cross-correlations as well as 
	the evolution of the structure over cosmic time scales
	\citep[e.g.,][]{Massey2007,Hirata2008,Mandelbaum2013,Kilbinger2013,Heymans2013,Planck2015lensing,Giannantonio2016}.
	These measurements can then also be used
	to test the laws of gravity as well, since growth of structure and the lensing effect itself depends on the nature 
	of gravity \citep[e.g.,][]{Simpson2013}. 

	\cite{Zhang2007} suggested the probe, $E_G$, as a consistency check on the theory of gravity by combining 
	RSD measurements with the galaxy-lensing cross correlations \citep[see also][]{Leonard2015}. 
	$E_G$ is sensitive to the ``gravitational slip'' 
	or the ratio of Newtonian potential and curvature potential, which are equal within GR in the absence of any 
	anisotropic stress. $E_G$ has been measured by \cite{Reyes2010}, \cite{Blake2016}, \cite{delaTorre2016}, \cite{Alam2016} and
	\cite{Amon2018} 
	using galaxy-galaxy lensing and by 
	\cite{Pullen2016} using galaxy-CMB lensing. These measurements are largely consistent with the \lcdm\ predictions, though 
	\cite{Pullen2016} measured \eg\ to be $\sim2.6\sigma$ lower than the predictions, with most of the discrepancy 
	coming from the low CMB lensing amplitude at large scales.
	
	In this work, we measure \eg\ using the BOSS galaxy samples and SDSS galaxy lensing as well as Planck CMB lensing 
	maps. In the case of galaxy lensing, due to limitations of the SDSS sample we only use the BOSS low redshift sample 
	(LOWZ) while for CMB lensing we use both LOWZ and CMASS samples.
		
	Throughout, we use the Planck 2015 cosmology \citep{Planck2015cosmo}, with $
	\Omega_m=0.309$, $n_s=0.967$, $A_s=2.142\times10^{-9}$, $\sigma_8=0.82$. 
	To get predictions for the matter correlation function, 
	we use the linear power spectrum with halofit
		\citep{Smith2003,Takahashi2012}, generated using the CAMB software \citep{Lewis2002}.

\section{Formalism and Estimators}\label{sec:formalism}
	In this section we provide a brief review of the theoretical formalism and the estimators used in this work.
	\subsection{Weak Lensing}\label{ssec:formalism_lensing}
		As photons travel from their sources to observers, their paths are deflected by the structure in the intervening matter 
		distribution \citep[see][for reviews]{Bartelmann2001,Kilbinger2015,2017arXiv171003235M}. 
		The lensing potential of a given lensing mass depends on 
		the lens potential and the geometric factors involving distances between the lens, source and observer, and is 
		given by
		\begin{equation}
			\Phi_L=\int \mathrm{d}\chi_l\frac{f_k(\chi_s-\chi_l)}{f_k(\chi_s)f_k(\chi_l)}\Psi(f_k
			(\chi_l)\vec\theta,\chi_l)
		\end{equation}
		$\chi_l$ and $\chi_s$ are line-of-sight distances to lens and source respectively ($\chi_s>\chi_l$), 
		$\vec\theta$ 
		is the angular separation between the lens and source on the sky 
          and $f_k(\chi)$ is 
          the transverse comoving 
		distance ($f_k(\chi)=\chi$ in a flat universe).
		 The Weyl potential $\Psi$ is given by
		\begin{equation}
			\Psi=\psi+\phi
		\end{equation}
		$\psi$ and $\phi$ are the Newtonian and curvature potentials. Within \lcdm, $\psi=\phi$ in the absence of 
		any anisotropic stress.
        The main focus of this paper is to test this equality of the two potentials through the measurement of the $E_G$ 
        parameter as defined in Section~\ref{ssec:formalism_EG}. 
	
		In the case where the angular extent of source is much smaller than angular scales over which lens potential 
		varies, the distortion matrix relating the source and observer planes is given by 
		\[
    		A=
	      \begin{bmatrix}
    	    1-\kappa-\gamma_1 & -\gamma_2\\
        	-\gamma_2 & 1-\kappa+\gamma_1
	      \end{bmatrix}.
  		\]
		where $\gamma=\gamma_1+i\gamma_2=|\gamma|e^{i2\phi}$ is the shear in the observer frame and can be rotated to 
		the lens-source frame to give $\gamma=\gamma_t+i\gamma_\times$.
		For a circularly symmetric lens, the convergence $\kappa$ and the tangential shear $\gamma_t$ are given by
	   \begin{align}
			&\kappa(r_p)=\frac{\Sigma(r_p)}{\Sigma_{c}}\\
			&\gamma_t(r_p)=\frac{\Delta\Sigma}{\Sigma_\text{crit}}=\frac{\overline{\Sigma}(<r_p)-\Sigma(r_p)}{\Sigma_{c}}
			\label{eq:gamma_t}
		\end{align}
 		while $\gamma_\times=0$ due to symmetry. 
		$\Sigma$ is the projected surface mass density and $\overline{\Sigma}(<r_p)$ is the mean $\Sigma$ within radius $r_p$ 
		from lens center. For non-circularly symmetric lens distributions, the equation is true when averaging within 
		annuli at fixed $r_p$.
		The geometric factor $\Sigma_\text{crit}$ is given by
		\begin{equation}\label{eq:sigma_crit}
    		\Sigma_\text{crit}=\frac{c^2}{4\pi G}\frac{f_k(\chi_s)}{(1+z_l) f_k(\chi_l) f_k(\chi_s-\chi_l)}.
		\end{equation}
		$1+z_l$ converts the $c^2/G$ factor to comoving space.
		The projected surface mass density can be written in terms of the 2-point galaxy-matter cross correlation 
		function
		(lensing is sensitive to density fluctuations, not the mean density)
		\begin{equation}\label{eq:losintgm}
    		\Sigma(r_p)=\bar\rho_m\int \mathrm{d}\Pi \,\xi_{gm}(r_p,\Pi)=\bar\rho_m w_{gm}(r_p),
		\end{equation}
		$\bar\rho_m$ is the mean matter density in comoving coordinates.
		Under the assumption of large projection length $\Pi$, the projected galaxy-matter cross correlation function 
		\wgm\ can be derived from the matter power spectrum as 
		\begin{align}
			w_{gm}(r_p)=b_g A_l \rcc\int \mathrm{d}z \,W(z)\int \frac{\mathrm{d}^2k}{(2\pi)^2}&P_{\delta
			\delta}(\vec{k},z)e^{i(\vec{r}_p\cdot\vec{k})}
		\end{align}
		$W(z)$ is the lens weight function, that we compute using the weights defined in 
		Sections~\ref{sssec:estimator_galaxy_galaxy_lensing} and~\ref{sssec:estimator_galaxy_cmb_lensing}. $b_g$ 
		is the galaxy bias, $r_{cc}$ is the galaxy-matter cross correlation 
		coefficient and $A_l$ is the scale independent lensing amplitude. Details of modeling the lensing measurements will be
		presented in a separate work (Singh et. al in prep). In this work we will derive the correction in \eg\ for the effects of 
		\rcc\ and non-linear galaxy bias from mocks as detailed in section~\ref{ssec:formalism_corrections} and 
		section~\ref{ssec:results_corrections}.
	
		\subsubsection{Estimator: Galaxy-galaxy lensing}\label{sssec:estimator_galaxy_galaxy_lensing}
			We measure the $\Delta\Sigma$ using tangential shear as
			\begin{equation}
    	    	\widehat{\Delta \Sigma}(r_p)=B_L(1+m_{\gamma})\frac{\sum_{ls}w_{ls}e_t^{(ls)}\Sigma_\text{crit}^{(ls)}}{2\mathcal R\sum_{rs}w_{rs}}-
				\frac{\sum_{rs}w_{rs}e_t^{(rs)}\Sigma_\text{crit}^{(rs)}}{2\mathcal R\sum_{rs}w_{rs}}
		       	\label{eq:delta_sigma_estimator}
    		\end{equation}
			where the summations are over all the lens-source (ls) or random lens-source (rs) pairs,
            \referee{where random lenses are unclustered random points with the same sky coverage
              and redshift distribution as the real lens galaxies}. 
			The signal measured around \refereee{random lenses} is 
			subtracted to remove the spurious signal from additive systematics \citep{Mandelbaum2005}.
			As demonstrated in \cite{Singh2016cov}, 
			subtraction of the signal around random points also leads to a more optimal estimator with better covariance 
			properties.
            Note that the 
			normalization factor uses weights computed using randoms. This is to account for the
            source galaxies that 
			are associated with the lens and are not lensed, but enter the estimator due to the scatter in photometric 
			redshifts. The lens-source pair weights are given by
			\begin{equation}\label{eq:delta_sigma_wt}
               w_{ls}=w_l\frac{\Sigma_\text{crit}^{-2}}{\sigma_\gamma^2+\sigma_{SN}^2}.
	    	\end{equation}
			$\Sigma_\text{crit}^{-2}$ enters because we have defined $\widehat{\Delta\Sigma}$ as the maximum likelihood estimator
			\citep{Sheldon2004}; $\sigma_{SN}$ is the shape noise and $\sigma_\gamma$ is the measurement noise. $w_l$ is the 
			weight for lens galaxies, defined in section~\ref{ssec:data_Boss}.
			
			As described in section~\ref{ssec:data_shapes}, we also multiply the ellipticity with shear responsivity $\mathcal R$ to get ensemble
			shear estimates and measured signal with the correction factors for
			photometric redshifts ($B_L\sim$1.1) and for the shear biases ($1+m_\gamma\sim$1.04).

		\subsubsection{Estimator: Galaxy-CMB lensing}\label{sssec:estimator_galaxy_cmb_lensing}
			Using CMB lensing, we measure the projected surface mass density as \citep{Singh2016} 
		    \begin{equation}
    	        \widehat{\Sigma}(r_p)=\frac{\sum_{lp}w_{lp}\kappa_{p}\Sigma_{c,{*}}}{\sum_{lp}w_{lp}}
				-\frac{\sum_{Rp}w_{Rp}\kappa_{p}\Sigma_{c,{*}}}{\sum_{Rp}w_{Rp}}
        	   \label{eq:sigma_cmb}
    		\end{equation}
			where the summation is over all the lens-pixel (pixels of CMB convergence map) pairs at separations 
			$r_p\in[r_{p,min},r_{p,max}]$ at the lens redshift and the signal around randoms is subtracted to remove 
			the effects of the correlated convergence, in the measurement \citep{Singh2016}. 

              The weight for each lens-pixel pair is given by
    		\begin{equation}
            	w_{lp}=w_l\Sigma_{c,*}^{-2}.
        	\end{equation}
			We have defined $\widehat{\Sigma}$ as the maximum likelihood estimator, 
			similar to galaxy-galaxy lensing, and $\Sigma_{c,*}$ is $\Sigma_\text{crit}$ with CMB as the source.

	\subsection{Projected galaxy clustering}\label{ssec:formalism_clustering}
		The two-point galaxy correlation function in redshift space can be written as
		\begin{align}
			\xi_{gg}(r_p,\Pi,z)=&b_g^2 \int \frac{\mathrm{d}^2k_\perp\mathrm{d}k_z}{(2\pi)^3}P_{\delta\delta}
			(\vec{k},z)\left(1+\beta\mu^2\right)^2e^{i(\vec k_\perp\cdot\vec r_p+k_z\Pi)}\label{eq:xi}
		\end{align}
		where $P_{\delta\delta}$ is the matter power spectrum; $r_p, k_\perp$ are the separation and
        Fourier vector in the plane of the sky while $
		\Pi,k_z$ are the separation and Fourier vector along the line-of-sight. 
		$b_g$ is the linear galaxy bias and the $(1+\beta\mu^2)$
		factor accounts for the linear redshift space distortions \citep[][see also Section
        \ref{ssec:formalism_rsd}]{Kaiser1987}, where $\beta=f(z)/b_g$ and $f(z)\approx
        \Omega_m(z)^{0.55}$ is the linear growth rate (effects of non-linear RSD on projected correlation functions are negligible and hence we 
        ignore them). 
		The projected correlation function is then
		\begin{align}
			w_{gg}(r_p)=&\int\mathrm{d}z \,W(z) \int_{-\Pi_\text{max}}^{\Pi_\text{max}}\mathrm{d}\Pi\, \xi_{gg}(r_p,
			\Pi,z)
			\label{eq:w}
		\end{align}
		$\Pi_\text{max}$ is the maximum line-of-sight separation for which the correlation function is computed.
		The weight function accounts for the differential contribution from different redshifts to the correlation
		function. It depends on total volume as well as redshift distribution
		of the galaxies, $p(z)$, and is
		given by \citep{Mandelbaum2011}
		\begin{equation}
			W(z)=\frac{p(z)^2}{\chi^2 (z)\mathrm{d}\chi/\mathrm{d}z} \left[\int \frac{p(z)^2}{\chi^2 (z)
			\mathrm{d}\chi/\mathrm{d}z} \mathrm{d}z\right]^{-1}
		\end{equation}
		Note that $W(z)$ can also be written in terms of comoving number density of galaxies, $n(z)$, as $W(z)\propto n(z)^2dV$.
		Finally the projected correlation function can be written as
		\begin{align}
			&w_{gg}(r_p)=\frac{b^2}{\pi^2}\int\mathrm{d}z W(z)\int_0^{\infty}\mathrm{d}k_z
			\int_0^{\infty}\mathrm{d}k_{\perp}\frac{k_\perp}{k_z} P_{\delta\delta}(\vec{k},z)\nonumber \\ &\sin(k_z\Pi_
			\text{max})J_0(k_\perp r_p)\left(1+\beta\mu^2\right)^2\label{eq:wgg}
		\end{align}
		where $J_0$ is the zeroth order Bessel function.
		
	\subsubsection{Estimator}\label{sssec:clustering_estimator}
		To measure the galaxy clustering, we use the generalized Landy-Szalay estimator \citep{Landy1993}
		\begin{equation}
				\xi_{gg}=\frac{(D-R)(D-R)}{RR}=\frac{DD-2DR+RR}{RR},
				\label{LSxi}
		\end{equation}
		where $D$ is the weighted galaxy sample, with weights defined in section~\ref{ssec:data_Boss} 
		and $R$ is for the random sample (corresponding to the weighted galaxy catalog).

		The projected correlation function can then be obtained by integrating the correlation function along the
		line-of-sight
		\begin{equation}
			w_{gg}=\int^{\Pi_{\text{max}}}_{-\Pi_{\text{max}}}\xi_{gg}(r_p, \Pi)\,\mathrm{d}\Pi.
		\end{equation}
		$\Pi_\text{max}$ denotes the size of top hat window function along the line-of-sight. It is desirable to
		 use a larger $\Pi_\text{max}$ to mitigate the effects of redshift space distortions in the projected correlation
		 function, though very large line-of-sight ($\Pi$) values add little signal and mostly contribute noise (for a 
		 survey of finite window size in redshift).
		 We use $\Pi_\text{max}=100\mpch$, though we also test our results using $\Pi_\text{max}=50\mpch$ and
		 $\Pi_\text{max}=200\mpch$, finding consistent results.

	\subsection{Redshift space distortions}\label{ssec:formalism_rsd}
	  The measured redshift ($z$) of the galaxy is a combination of Hubble recession
      and the peculiar velocity\footnote{There are also small higher order contributions from various relativistic effects 
       \citep{Cappi1995, Alam2017MNRAS.470.2822A}, which we will ignore.}. 
      The peculiar velocity component of the redshift affects the line-of-sight distance to a galaxy determined using 
      the cosmological distance-redshift relation, introducing an anisotropy in the two-point 
      correlation function. Within linear theory, the real space power spectra can be converted to {redshift} 
      space power spectra as \citep{Kaiser1987} 
      \begin{align}
			&P_{gg}^s(\vec k)=P_{gg}^r(\vec k)(1+\mu_k^2\beta)^2.
      \end{align}
      where $P_{gg}^r$ is the real space galaxy power spectrum, $P_{gg}^s$ is the redshift space galaxy power 
      spectrum, $\mu$ is the cosine of the angle made by the wave-vector $\vec k$ from the line-of-sight and $\beta$ is 
      the RSD parameter. \cite{Hamilton1992} extended this approach to the two-point correlation function and showed 
      that the linear theory two-point correlation function in redshift space is given by
      \begin{equation}
			\xi_{gg}(\vec s)=\left[1+f(\partial/\partial z)^2 (\nabla^2)^{-1}\right]^2\xi_{gg}(\vec r)
      \end{equation}
	   where $\xi_{gg}(\vec r)$ and $\xi_{gg}(\vec s)$ are the galaxy correlation function in real and redshift space. 
	   The development of perturbation theories has produced better models of redshift
       space correlation that work even in the quasi-linear scales \citep{2002PhR...367....1B, 2006PhRvD..73f3519C, 2012PhRvD..85h3509C, 2012MNRAS.427.2537C}.  
       They enable interpretation of RSD measurements to
       smaller separations, and hence result in  more precise measurements of RSD \citep{2014MNRAS.439.3504S}. 
       {There are also efforts to model non-linear scales using HOD \citet{2014MNRAS.444..476R}, by combining N-body simulations with Eulerian 
       perturbation theory \citet{2017JCAP...10..009H} and using higher order LPT calibrated with N-body simulations \citet{2018arXiv180104950S}.}
	   We use Convolution 
	   Lagrangian Perturbation Theory (CLPT) to predict the real space correlation function and velocity moments, which 
	   are then combined with the Gaussian Streaming Model (GSM) to predict the redshift-space correlation function 
	   \citep{Carlson2012,Wang2013}. We use tools developed in \cite{AlamRSD2015} which have been tested on several 
	   mocks and N-body simulations including the completed cosmological analysis of DR12 from BOSS  
	   \citep{2016arXiv160703148S, Boss2016combined}. 
	   {Our formalism ignores the impact of gravitational lensing and generalized  Sachs-Wolfe effects on the two point clustering which are shown 
	   to be negligible at the level of current precision \citep{2009PhRvD..79b3517Y}. }

		\subsubsection{Measuring the growth rate}
     		We estimate $\beta$ using the monopole and quadruple moments of the galaxy auto-correlation function, 
		 	obtained by projecting the redshift space correlation function onto the Legendre basis.
			\begin{equation}
				\xi_{gg,l} (s)=\int_{-1}^1\mathrm{d}\mu\, \mathcal{P}_l(\mu) \xi_{gg}(s,\mu)
			\end{equation}
			where $s$ is the distance in redshift space, $\mu$ is the angle between $\vec{s}$ and the plane of the sky 
			and $\mathcal P_l$ is the Legendre polynomial of order $l$ ($l=0$ 
			for monopole and $l=2$ for quadrupole).
			
			We  use COSMOMC \citep{cosmomc} to run a Markov Chain Monte Carlo (MCMC) fit for the multipole moments of 
			the 
			correlation function. The correlation function is fit for 3 parameters ($f,\sigma_{FOG},\nu$), where 
			$f$ is the growth rate, $\sigma_{FOG}$ is an additional velocity dispersion to model the Finger of God 
			effect 
			\citep{2012MNRAS.426.2719R,AlamRSD2015} and $\nu$ is the over-density which determines the first and second
            order bias through the peak-background split \citep{White2014}:
		      \begin{align}
        		 F^\prime &= \frac{1}{\delta_c} \left[ a\nu^2-1 +\frac{2p}{1+(a\nu^2)^p}\right] \\
		         F^{\prime\prime}&= \frac{1}{\delta_c^2} \left[ a^2\nu^4-3a\nu^2 +\frac{2p(2a\nu^2+2p-1)}{1+(a\nu^2)^p} 
		         \right]
		      \end{align}
		    where a = 0.707, p = 0.3 gives the Sheth-Tormen mass function \citep{ST1999}, and $\delta_c=1.686$ is the 
		    critical density for collapse. $F^\prime$ and $F^{\prime\prime}$ are the first and second order Lagrangian 
		    bias. 
		    The linear galaxy bias is given by $b_g=1+F^\prime$. The RSD fits use scales from \referee{35~to }
		    70~$h^{-1}$Mpc. The minimum scale is chosen to only use the scales where the model is shown to be consistent 
		    with survey mocks and N-body simulations \citep{AlamRSD2015}, whereas the maximum scale is determined by the 
		    size of the jackknife regions that we use to estimate the covariance (see section~\ref{ssec:formalism_covariance}).
			
	\subsection{$\Upsilon$ Estimator}\label{ssec:formalism_upsilon}
		In galaxy-galaxy lensing, we measure $\Delta\Sigma=\overline{\Sigma}(<r_p)-\Sigma(r_p)$. 
		$\overline{\Sigma}(<r_p)$ and hence $\Delta\Sigma$ is affected by all scales $<r_p$.
		This induces considerable theoretical uncertainty in modeling the 
		$\Delta\Sigma$ observable due to the lack of a good model for the small scales, for example,  non-linear 
		clustering and galaxy-dark matter cross-correlations (usually quantified through the $\rcc$ parameter) at
		scales comparable to or within the virial radii of dark matter halos. To mitigate the impact
        of small-scale theoretical uncertainty on large-scale structure measurements,
        \cite{Baldauf2010} suggested a new estimator 
		\begin{equation}\label{eq:upsilon_gm}
			\Upsilon_{gm}(r_p,r_0)=\Delta\Sigma_{gm}(r_p)-\frac{r_0^2}{r_p^2}\Delta\Sigma(r_0)
		\end{equation}
		Expanding $\Delta\Sigma$ in terms of the correlation function, it can be shown that $\Upsilon$ is independent of
		information from small scales, $r<r_0$:
			\begin{multline}
			\Upsilon_{gm}(r_p,r_0)=\bar{\rho}_M\left[\frac{2}{r_p^2}\int_{r_0}^{r_p} \mathrm{d}r'\, w_{gm}(r')-  \right. \\
			\left. w_{gm}(r_p)+\frac{r_0^2}{r_p^2}w_{gm}(r_0) \right].
			\label{eq:upsilon_gm_long}
		\end{multline}

		Analogous to this lensing observable, we can define $\Upsilon_{gg}$, starting by defining
		\begin{equation}
			\Sigma_{gg}(r_p)={\rho}_\text{crit}\int \mathrm{d}\Pi\, \xi_{gg}(r_p,\Pi),
		\end{equation}
		where we have ignored the effects of the mean density and
		$\rho_\text{crit}$ is used to get $\Sigma_{gg}$ in units of density. The choice of $\rho_\text{crit}$ is also
		useful for the definition of the $E_G$ parameter in Section~\ref{ssec:formalism_EG}. The definition of
		$\Upsilon_{gg}$ follows from the definition of $\dsigma$ and Eq.~\eqref{eq:upsilon_gm}.

	\subsection{$E_G$ Statistic}\label{ssec:formalism_EG}
		The perturbed metric in the conformal Newtonian gauge is written as
		\begin{equation}
			ds^2=a(\tau)^2\{-(1+2\psi)d\tau^2+(1-2\phi)(dr^2+r^2d\Omega^2) \}
		\end{equation}
			As stated in section~\ref{ssec:formalism_lensing}, $\psi=\phi$ in \lcdm, in the absence of
			anisotropic stresses. 
              However, several modified gravity theories lead to modifications in
			either or both the potentials and in general violate the equality of two potentials \citep{Jain2010}.

			\cite{Zhang2007} proposed an estimator, $E_G$, to test the equality of two metric potentials
		\begin{equation}
			E_G(k,z)=\frac{P_{g\kappa}(k,z)}{P_{g\theta}(k,z)}
		\end{equation}
			The projected real-space analog of $E_G$ was defined by \cite{Reyes2010} (see also \citealt{Leonard2015}) as
		\begin{equation}
			E_G(r_p)=\frac{\Upsilon_{gm}(r_p)}{\beta\Upsilon_{gg}(r_p)}
		\end{equation}
		Equivalently, we can define \eg\ in terms of projected surface mass density
		\begin{equation}
			E_G(r_p)=\frac{\Sigma_{gm}(r_p)}{\beta\Sigma_{gg}(r_p)}
		\end{equation}
		In the case of galaxy-galaxy lensing, we will measure \eg\ using $\Upsilon$ estimators, while for CMB lensing
		we will present results using both $\Upsilon$ and $\Sigma$. $\Sigma$ provides a more accurate and precise
        measurement at small scales 
		as there is no mixing of scales or removal of information as in $\Delta\Sigma$ or $\Upsilon$. However, $\Sigma$ 
		has lower signal to noise ($S/N$) at large scales due to the low amplitude and higher contribution from cosmic 
		variance, which
		introduces a noise bias in $\eg$ (because we take the ratio of quantities with low $S/N$), 
		in addition to the measurement being 
		noisy. $\Upsilon$ on the other hand has 
		lower $S/N$ at small scales but is better at large scales due to higher $S/N$.
		
		$E_G$ is sensitive to the difference between $\psi,\phi$ as $P_{g\kappa}$ or $\Upsilon_{gm}$, measured from
		lensing, are sensitive to the Weyl potential ($\Psi=\psi+\phi$), while the linear growth rate
		of matter perturbations depends on the Newtonian potential $\psi$.
		In $\Lambda$CDM, with $\psi=\phi$
		\begin{equation}
			E_G(r_p,z)=\frac{\Omega_m(z=0)}{f(z)}
		\end{equation}
		In this paper we focus on measurements of \eg\ and their implications for the \lcdm\ model. We do not test for 
		any specific models of modified gravity, and refer the reader to \cite{Pullen2015a} for $E_G$
        predictions in some alternate theories of gravity.

	\subsection{\eg\ corrections}\label{ssec:formalism_corrections}

		Theoretically, in the linear regime, $E_G$ is a clean null test for the equality of the two metric potentials 
		and hence a test of GR.
		However, it can be subject to some of the systematics that we discuss here.

		\emph{Non-linear clustering} at small scales can lead to significant deviations from the expected value of 
		$E_G$,
		especially since it affects lensing and clustering measurements differently. The clustering ($\Upsilon_{gg}$)
		amplitude goes as $b_g^2$, which includes difficult-to-model non-linear galaxy bias. Lensing
        ($\Upsilon_{gm}$) is sensitive to $b_g r_{cc}$, where $b_g$ includes non-linear bias while \rcc\ is the
		galaxy-matter cross correlation coefficient. $\rcc\sim 1$ at large scales but deviates strongly at small
		scales and depends on details of how galaxies populate halos
		\citep[see e.g.][for more detailed discussion on $r_{cc}$]{Baldauf2010}. As discussed in
		\cite{Baldauf2010}, use of $\Upsilon$ rather than $\Delta\Sigma$ partially mitigates the effects of non-linear 
		clustering by removing small-scale information, though
		the efficiency of this small-scale removal depends on the choice of $r_0$. The choice of $r_0$ depends on the 
		balance between theoretical
		uncertainties, which are large at small scales, and statistical errors, which increase 
		with $r_0$ as more signal is removed.
		 For our main analysis, we choose $r_0=2\mpch> 2r_\text{vir}$, as suggested by \cite{Baldauf2010}. 
		 This still leaves some residual effects
		 from non-linear scales and we attempt to mitigate these effects by estimating them using simulations. For this 
		 we compute the correction factor $C_{nl}$, in real space, 
		 to correct for the effects of scale-dependent bias and $r_{cc}$.
		\begin{equation}\label{eq:Cnl}
			C_{nl}(r_p)=\frac{\ugg^\text{sim}(r_p)}{b_\text{linear}\ugm^\text{sim}(r_p)}
		\end{equation}
		$b_\text{linear}$ is the linear galaxy bias. 

		The finite size of the top-hat window function in the clustering measurement leaves some residual effects of
		\emph{linear redshift space distortions} in the projected correlation function (see also
		Section~\ref{sssec:clustering_estimator}) and hence in the $E_G$ measurement.
		The lensing measurement is not affected by RSD to first order and also the implicit line-of-sight integration in lensing 
		is 
		much longer ($\Pi_\text{max}^\text{lens} \gg 100\mpch$). 
      We mitigate the effect of the limited window in 
		projected clustering using the corrections computed with the linear theory and Kaiser formula 
		\citep{Kaiser1987}.
		\begin{equation}
			C_\text{rsd+win}(r_p)=\frac{\wgg(r_p|\Pi_\text{max}=100\mpch)}{\wgg(r_p|\Pi_\text{max}=\infty)}
		\end{equation}
		where \wgg\ is defined in Eq.~\eqref{eq:wgg}.

		In addition to the above, the window function for lensing (depends on the broad
		lensing kernel and systematics weights) also varies, due to which effective weights to different $\Pi$
		bins are different for
		lensing and clustering. We correct for this effect using simulations by computing $\Upsilon_{gm}$ separately with top hat weights
		and lensing weights which also includes the lens-source pair weights defined in 
		Section~\ref{ssec:formalism_lensing}. 
		\begin{equation}
			C_\text{lens win}(r_p)=\frac{\int d\Pi \,
			\xi_{gm}(r_p,\Pi)}{\int d\Pi \,\mathcal W_{gm}(\Pi)\,\xi_{gm}(r_p,\Pi)}
		\end{equation}
		Here $\mathcal W_{gm}$ is the lensing window function
		derived in Appendix~\ref{append:lensing_window}.

		Due to different redshift weights for lensing and clustering, the two measurements are also at different 
		effective redshifts. 
		Since the lensing amplitude scales as $b_gD(z)^2$ ($D(z)$ is the linear growth function) 
		and clustering amplitude scales as
		$b_g^2D(z)^2$, we correct for the different effective redshifts as
		\begin{equation}
			C_{z}(r_p)=
			\frac{D(z=z_\text{eff}^{gg})}{D(z=z_\text{eff}^{gm})}\sqrt{\frac{\int dz \mathcal{W}_{gg}(z) 
			\wgg(r_p,z)}{\int 
			dz \mathcal{W}_{gm}(z) \wgg(r_p,z)}}
		\end{equation}
		The ratio is directly computed from the data by measuring clustering with lensing weights 		
		($\mathcal W_{gm}$) assigned to the galaxies and is in general scale-dependent as different scales grow differently with redshift under the 
		effects of non-linear physics. The ratio of growth function is computed from the theoretical model at fiducial 
		cosmology. Within linear theory, this correction leads to cancellation the redshift dependent factors from projected clustering and lensing 
		and the final measurement is at the effective redshift of measured growth rate $f$, which has the effective redshift of 
		clustering.

		There are also additional number density fluctuations due to the lensing effects of the intervening 
		large scale structure between the observer and galaxies \citep{Dizgah2016}. We estimate the impact of these 
		effects in appendix~\ref{append:lensing_EG} and apply a correction $C_\text{lens}$.

		We estimate and show these corrections in section~\ref{ssec:results_corrections}.
		The final correction applied to the measured $E_G$ signal is the product of all corrections defined above.
		\begin{equation}
			C_\text{tot}(r_p)=C_\text{lens win}(r_p)\times C_{nl}(r_p)\times C_\text{rsd+win}(r_p)\times C_z(r_p)
								\times C_\text{lens}
		\end{equation}
	
	     \subsection{Covariance}\label{ssec:formalism_covariance}
		 To compute the covariance matrix for both clustering and lensing measurements, we split the sample into 100 jackknife 
		 regions with approximately equal area on the sky.  
         \cite{Singh2016cov} showed that the jackknife errors are consistent with
		 theoretical expectations when using mean zero quantities, i.e., galaxy field with mean subtracted using 
		 randoms, as for the lensing and clustering estimators in this work. 
		 Throughout, the quoted errors and uncertainties will be jackknife errors. 
		 Whenever fitting models or computing derived quantities such as \eg\ (or a mean), we do so for each
		 jackknife region separately using \referee{the diagonal entries of the jackknife covariance} 
		 and then quote the jackknife mean and errors of model
         parameters.  Lensing measurements 
         are dominated by shape noise on the scales that contribute the most information, so using diagonal covariance for lensing and \eg\ 
         is justified (covariance of \eg\ is dominated by lensing). However, the use of a diagonal covariance matrix 
         can lead to biased parameter and error estimation when the bins are strongly correlated due to incorrect weighting applied to the 
         bins. Similar biases can also arise when inverting
         noisy covariance matrices, which are accounted for (but not corrected for) by Hartlap factor \citep{Hartlap2007,Taylor2013} by 
         increasing the size of errors.
         In our work this issue is most relevant for the measurement of the RSD parameters from the
         clustering measurement, 
         where the differences in the mean value from using the full versus diagonal \referee{jackknife} covariance can be $\sim0.5\sigma$ 
         and the errors when using the diagonal covariance are also overestimated by $\sim10\%$.
         Thus for RSD, we fit each jackknife region separately, but using the full covariance matrix estimated using all regions. 
         Also, for the errors quoted on the parameters, we do not apply the Hartlap factor, which is close to 0.84 for our analysis and will
         increase the quoted errors by $\sim 8\%$. Since for deriving \mean{\eg} 
         we use only the diagonal covariance with the average over the jackknife, it is 
         not clear whether Hartlap factor is the right correction to use and hence we do not apply it to our results.
         \referee{If we make a simplifying assumption that the uncertainty in each bin is estimated independently of the other bins, then the Hartlap 
         factor in each bin is the same as that for the case with one data point; this correction factor is $\sim1.5\%$ for 100 
         jackknife regions.} 

\section{Data}\label{sec:Data}
	We use the same datasets as used in \cite{Singh2016}. In this section, we briefly describe these datasets for 
	completeness.
   \subsection{SDSS}
	\referee{We use imaging and spectroscopic datasets from the SDSS-I,II and III surveys \citep{1998AJ....116.3040G,2000AJ....120.1579Y,
	2001AJ....122.2267E,2002AJ....123.2945R,2002AJ....124.1810S,Gunn2006}. The SDSS-I/II survey was completed in the seventh data release
    	\citep{2009ApJS..182..543A} though the data reduction pipeline was improved by \citep{2008ApJ...674.1217P} in the eighth data
    	release \citep{2011ApJS..193...29A}, which is used to derive our imaging catalogs as
        described in the next section.}

	\subsection{Galaxy lensing: Shape sample}\label{ssec:data_shapes}
		The catalogue of galaxies with measured shapes used in this
		paper (described in \citealt{Reyes2012} and further characterized in \citealt{Mandelbaum2013})
		was generated using
		the re-Gaussianization method \citep{2003MNRAS.343..459H} of
		correcting for the effects of the point-spread function (PSF) on the
		observed galaxy shapes. The catalogue production procedure was described in detail in previous work, so
		we describe it only briefly here.  Galaxies were selected in a 9243
		deg$^2$ region, with an average number density of $1.2$ arcmin$^{-2}$ based on requiring
        shape measurements in both $r$ and $i$ bands.
		The selection was based on cuts on the imaging quality, data reduction
		quality, galactic extinction $A_r<0.2$ defined using the dust maps from
		\cite{1998ApJ...500..525S} and the extinction-to-reddening ratios from
		\cite{2002AJ....123..485S}, apparent magnitude (extinction-corrected
		$r<21.8$ using model magnitudes\footnote{\texttt{http://www.sdss3.org/dr8/algorithms/\\magnitudes.php\#mag\_model}}),  and 
		galaxy size compared to the
		PSF.  For
		comparing
		the galaxy size to that of the PSF, we use the resolution factor $R_2$
		which is defined using the trace of the moment matrix of the PSF
		$T_\mathrm{P}$ and of the observed (PSF-convolved) galaxy image
		$T_\mathrm{I}$ as
		\beq
		R_2 = 1 - \frac{T_\mathrm{P}}{T_\mathrm{I}}.
		\eeq
		We require $R_2>1/3$ in both $r$ and $i$ bands.

		The basic principle of shear measurement
		using these images is to fit a Gaussian profile with elliptical
		isophotes
		to the image, and define the components of the ellipticity
		\beq
		(e_+,e_\times) = \frac{1-(b/a)^2}{1+(b/a)^2}(\cos 2\phi, \sin 2\phi),
		\label{eq:e}
		\eeq
		where $b/a$ is the axis ratio and $\phi$ is the position angle of the
		major axis.  The ensemble average ellipticity is then an estimator for the shear,
		\beq
		(\gamma_+,\gamma_\times) = \frac{1}{2\cal R}
		\langle(e_+,e_\times)\rangle,
		\eeq
		where ${\cal R}\approx 0.87$ is called the `shear responsivity' and
		represents the response of the distortion to a small
		shear \citep{1995ApJ...449..460K, 2002AJ....123..583B}; ${\cal R} \approx
		1-e_\mathrm{rms}^2$.

		For this work, we do not use the entire source catalogue, only the
		portion overlapping the LOWZ sample.

		When computing the weak lensing signals around the
		LOWZ galaxies, we need estimates of the redshifts for the fainter
		source galaxies. For this purpose, we
		use the maximum-likelihood estimates of photometric redshifts (photo-$z$) based on the five-band
		photometry from the Zurich Extragalactic
		Bayesian Redshift Analyzer \citep[ZEBRA,][]{2006MNRAS.372..565F}, which were
		characterized by \cite{Nakajima2012} and \cite{Reyes2012}.  
		Following \cite{Nakajima2012}, we use a representative calibration sample of source galaxies with spectroscopic
		redshifts to calculate biases
		in weak lensing signals due to bias and scatter in the photo-$z$, and
		applied corrections that were of order 10 per cent ($\pm 2$ per cent)
		to the weak lensing signals (we multiply the final lensing measurement with 1.1 and the 2 percent uncertainty is added to 
		the systematics error budget). 
		We further test the accuracy of these corrections using the clustering-redshift method \citep[e.g.,][]{Menard2013} 
		in appendix~\ref{append:clustering_pz}.

As discussed in \cite{2017arXiv171000885M}, new shear calibration simulations that include the
impact of nearby neighbors on the shear estimates from re-Gaussianization identified a previously
uncorrected effect due to those neighbors.  Since the shear calibration simulations that were
originally used to quantify shear systematic errors in 
the SDSS catalog used in this work \citep{2012MNRAS.420.1518M} did not include nearby neighbors, we
must include this newly-identified correction in our results in this work.  Figure 18 in
\cite{2017arXiv171000885M} clearly illustrates that this correction is a function of the depth of
the sample used for shear estimation.  For SDSS-like depths, the correction is a factor of $1.04\pm
0.02$.  We apply the 1.04 correction to our LOWZ lensing measurements, and fold the $\pm 0.02$ into
our systematic error budget.

	\subsection{SDSS-III BOSS}\label{ssec:data_Boss}
		\referee{The BOSS spectroscopic survey was performed using the BOSS spectrograph \citep{Ahn:2012,Smee:2013} with  targets selected from the 
		SDSS photometric catalogs \citep{Dawson:2013} and data processed by automated pipelines \citep{Blanton:2003,Bolton:2012}.
		We use SDSS-III BOSS data release 12 \citep[][]{Alam2015, Reid2016} and select 
		LOWZ galaxies in the redshift range $0.16<z<0.36$ and CMASS galaxies in $0.45<z<0.7$.}
		
		\referee{
		The LOWZ sample consists of Luminous Red Galaxies (LRGs) at $z<0.4$ and  is approximately volume-limited in the
		redshift range we use, $0.16<z<0.36$, with a number
		density of $\bar{n}\sim 3\times10^{-4}~h^3\text{Mpc}^{-3}$ \citep{Manera2015,Reid2016}. 
		}
		To test for the redshift 
		evolution of $E_G$, we also split the sample into 
		two redshift bins, Z1: $0.16<z<0.26$ and Z2: $0.26<z<0.36$. Further, we also use a sample of field galaxies 
		\citep{Singh2016} defined using the counts-in-cylinders (CiC) technique \citep{Reid2009}. Field galaxies are mainly 
		central galaxies and have somewhat lower bias and hence are relatively less affected by
        non-linear bias effects.
		
		The BOSS CMASS sample consists of higher redshift galaxies ($0.45< z < 0.7$) targeted using color and magnitude 
		cuts intended to select a uniform sample of massive galaxies \citep{Reid2016}. 
		The use of apparent color and magnitude to select targets may affect the selected galaxy
        sample due to relativistic effects. \cite{Alam2017MNRAS.471.2077A} showed that impact of
        such effects are negligible for the CMASS sample.
		
		We also use the weights defined by the BOSS collaboration \citep{Ross2012} for systematics ($w_\text{sys}$), 
		redshift 
		failures ($w_\text{no-z}$) and fiber collisions ($w_\text{cp}$). The effects of these weights for both 
		clustering 
		and lensing measurements were studied in detail in \cite{Singh2016} \citep[see also][]{Ross2012, Anderson2014}. 
		The final weights are defined as
		\begin{equation}
			w=w_\text{sys}(w_\text{no-z}+w_\text{cp}-1)
		\end{equation}

	\subsection{Planck CMB lensing maps}
		For CMB lensing, we use the publicly available lensing maps from Planck collaboration \citep{Planck2015lensing}. 
		We convert the provided convergence values in Fourier space $\kappa_{\ell,m}$ into real space 
		$\kappa_{\theta,\phi}$
		using {\sc healpy} \citep{Gorski2005} with $n_\text{side}=1024$. We use the full $\ell$ range ($8<\ell<2048$) 
		when 
		constructing the map. \cite{Planck2015lensing} found some evidence of systematics in lensing maps and
		use $40<\ell<400$ for the main cosmological analysis. However, in \cite{Singh2016} no evidence of systematics 
		was found when cross-correlating the full convergence map with the BOSS galaxies and hence we will use the full 
		$\ell$ range in this work as well, \referee{though we will quote the number using scales corresponding to $\ell<400$ cut}. 
		We refer the reader to \cite{Singh2016} for more details about choice of 
		pixel size in the maps and various systematic tests that were carried out.
		
	\subsection{Simulations}\label{ssec:data_simulations}

		To test our \eg\ pipeline and compute the corrections described in section~\ref{ssec:formalism_corrections},
		we use the `Med-Res' simulations described in \cite{Reid2014}, with snapshots at $z=0.25$ 
		and $z=0.40$ for the LOWZ sample and $z=0.60$ for the CMASS sample. The sample of galaxies is 
		generated 
		using the HOD model from \cite{Zheng2005} to fit the clustering of galaxies for full LOWZ or CMASS samples,
		 assuming a fixed abundance of halos 
		\citep[see][for more details]{Reid2014}. 
		We will use same simulation catalogs to compute corrections for the 
		subsamples as well.
		To compute the galaxy-matter cross-correlations, we cross-correlate 
		the galaxies with the randomly subsampled matter particles. 

\section{Results}\label{sec:results}
	In this section we presents the results, starting with 
	growth rate measurements in the data, then \eg\ measurements in simulations and tests for
    systematics and then
	the $E_G$ measurements in the data. 

      \begin{table}
			\begin{tabular}{|c|c|c|c|c|}
                \hline
Sample & Lensing & $z_\text{eff}$ & $f\frac{\sigma_8}{\sigma_{8,fid}}$  & $E_G$\\ 
& map& & &\\ \hline 
Z1  &SDSS  &$0.21$ &$0.75_{-0.088}^{+0.097}$ [$0.64$] &$0.38_{-0.061}^{+0.052}$ [$0.48$]  \\ \hline 
Z1  &Planck  &$0.21$ &$0.75_{-0.088}^{+0.097}$ [$0.64$] &$0.6_{-0.113}^{+0.112}$ [$0.48$]  \\ \hline 
Lowz  &SDSS  &$0.27$ &$0.62_{-0.056}^{+0.055}$ [$0.66$] &$0.4_{-0.038}^{+0.046}$ [$0.46$]  \\ \hline 
lowz-pl  &Planck  &$0.27$ &$0.62_{-0.056}^{+0.055}$ [$0.66$] &$0.46_{-0.088}^{+0.082}$ [$0.46$]  \\ \hline 
Field  &SDSS  &$0.27$ &$0.62_{-0.056}^{+0.055}$ [$0.66$] &$0.4_{-0.039}^{+0.042}$ [$0.46$]  \\ \hline 
Field  &Planck  &$0.27$ &$0.62_{-0.056}^{+0.055}$ [$0.66$] &$0.47_{-0.079}^{+0.077}$ [$0.46$]  \\ \hline 
Z2  &SDSS  &$0.32$ &$0.5_{-0.065}^{+0.063}$ [$0.69$] &$0.45_{-0.073}^{+0.08}$ [$0.45$]  \\ \hline 
Z2  &Planck  &$0.32$ &$0.5_{-0.065}^{+0.063}$ [$0.69$] &$0.4_{-0.1}^{+0.115}$ [$0.45$]  \\ \hline 
CMASS  &Planck  &$0.57$ &$0.69_{-0.07}^{+0.066}$ [$0.77$] &$0.4_{-0.051}^{+0.049}$ [$0.4$]  \\ \hline 

\referee{CMASS}  &Planck & & $r_p\in[25,150]$ &$0.38_{-0.064}^{+0.065}$ [$0.4$]  \\ \hline 
  		\end{tabular}
			\caption{ Values of the growth function, $f$, and \eg\ 
				for different samples and using SDSS and Planck CMB lensing maps. 
			The values in square brackets, $[]$, are the predicted 		
			values from the Planck \lcdm\ model. $z_\text{eff}$ is the effective redshift of galaxy clustering measurements (Galaxy-lensing cross 
			correlations have different $z_\text{eff}$ for which we apply $C_z$ correction to \eg).
			The errorbars in this table are statistical only. 
			\refereee{There is additional systematic uncertainty of order $\sim3\%$ in measurements using Planck CMB lensing maps and 
			$\sim6\%$ in measurements using SDSS galaxy lensing. The systematic uncertainties include contributions 
			 of order $2\%$ from the applied corrections} \referee{ and $\sim2\%$ on RSD modeling 
			\citep{Alam2016} (modeling uncertainties). In addition, 
			measurements using SDSS galaxy lensing 
			have a further $5-6\%$ systematic uncertainty from shear calibrations ($\sim2\%$) and photometric redshifts ($\sim 5\%$).}
			\refereee{We do not estimate observational systematic uncertainties in CMB lensing maps.}
			Note that for the 
			field sample, we use $f$ measured from the full LOWZ sample. Due to the selection effects from CiC method, 
			$f$ from the field sample is somewhat biased ($f=0.74\pm0.05$).
			\referee{In the last row we have quoted the \eg\ measurement using scales
              $25<r_p<150$\mpch, which corresponds approximately to 
			the range of scales ($40<\ell<400$) used by \citealt{Pullen2016} in their \eg\ measurement.}
			}
			\label{tab:params}
		\end{table}

	\subsection{Growth rate measurement}\label{ssec:results_growth_rate}
	   \input{GrowthM}
	   	   
	   \begin{figure}
	         \centering
	         \includegraphics[width=0.9\columnwidth]{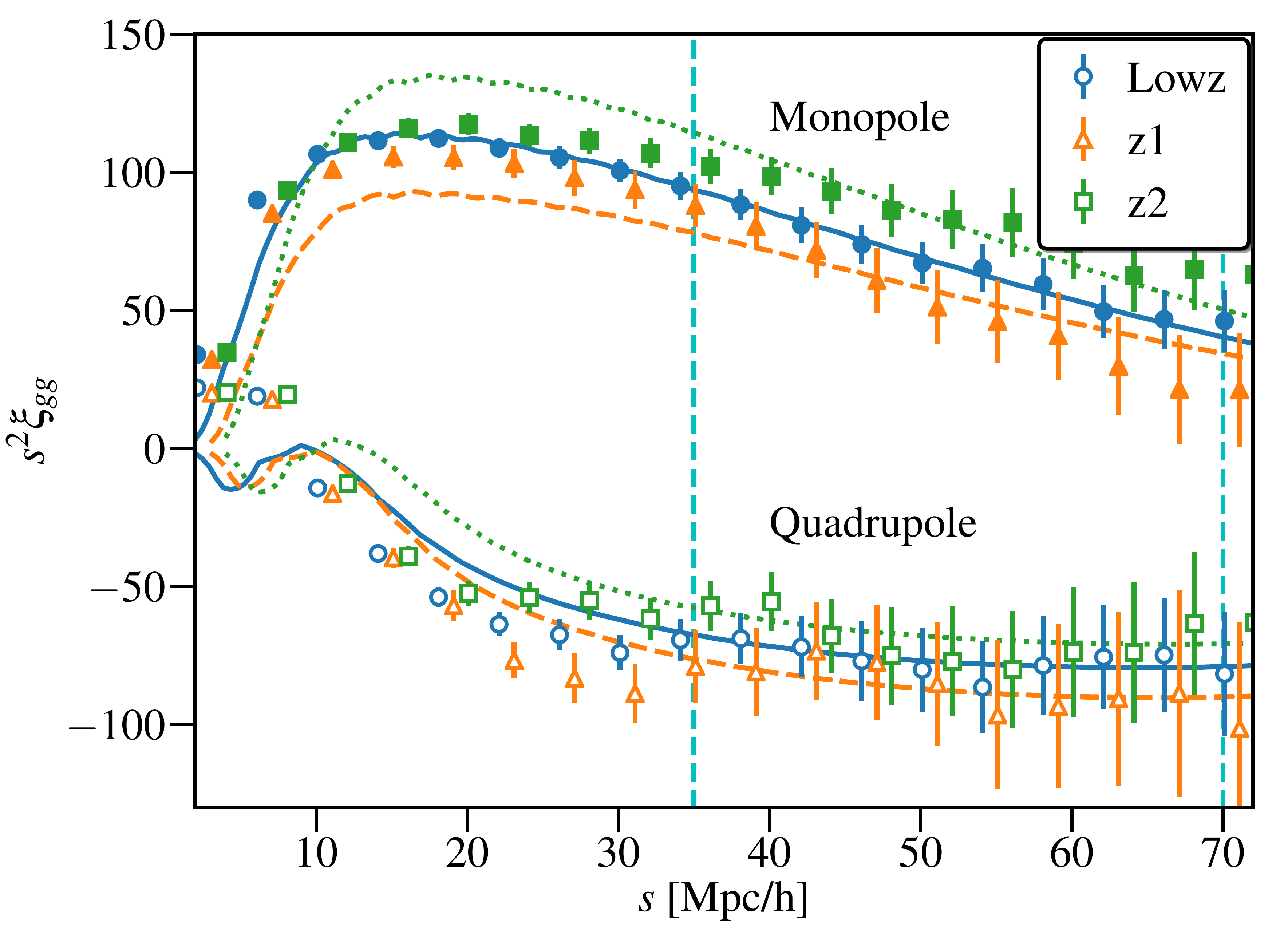}
		     \caption{For different subsamples of the LOWZ sample in the two panels, we show the
           		multipole moments -- the monopole ($l=0$) and quadrupole ($l=2$) 
	     		of the galaxy correlation function, along with the best-fitting model (solid lines). The vertical dashed lines mark the range
				within which the model was fit, \referee{$35<r_p<70\mpch$}
	     }
	     \label{fig:rsd_multipole}
      \end{figure}

	\subsection{$E_G$ in simulations and corrections}\label{ssec:results_corrections}
\begin{figure}
        	\centering
         	\includegraphics[width=\columnwidth]{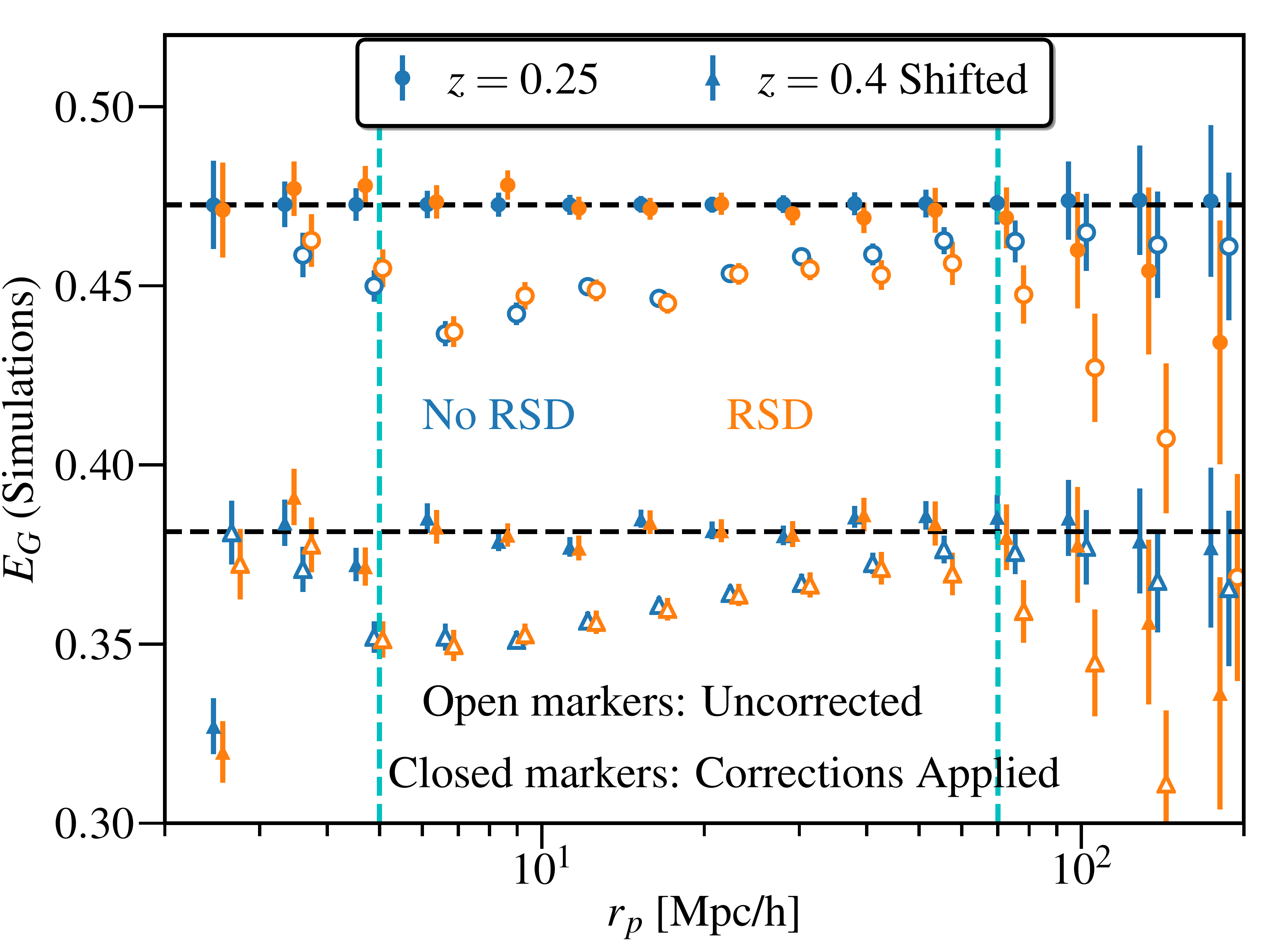}
	         \caption{\eg\ measurements using simulations. Points with
		         errorbars show the measurements, while dashed black lines show \lcdm\
		         predictions at two redshifts. Vertical cyan lines mark the scales $5<r_p<70\mpch$ within which 
		         we measure the $\mean{\eg}$ (not shown).  
                 Blue points show
	        	 the measurements using real space positions for galaxies while orange 
		         points show measurements in redshift space. Open markers show
		         measurements without any corrections applied while closed markers show
	    	     results obtained after applying all corrections. 
 					We find that the most important
		         correction is for the combined effects of non-linear bias and $\rcc$ (see eq.~\eqref{eq:Cnl}). 
                 The corrections terms for the LOWZ sample are shown in 
	    	     Fig.~\ref{fig:EG_corrections}. 
	          }
    	     \label{fig:EG_sims}
        \end{figure}

		In Fig.~\ref{fig:EG_sims} we show the $E_G$ measurements in the simulation box with snapshots at
		$z=0.25$ and $z=0.4$, in both real space and redshift space. Without 
		applying any corrections, \eg\ is biased low with considerable scale dependence. The average \eg, \mean{\eg}, over all 
		scales 
		is low by $\sim5\%$ compared to the \lcdm\ predictions both in real space and redshift space. RSD effects are
		important at large scales, but due to relatively lower $S/N$ these scales do not contribute much
		when computing the mean \eg. Still we do correct for these effects as described in 
		Section~\ref{ssec:formalism_corrections}. 
		The biggest contribution to the bias in \eg\ is from the combined effects
		of non-linear galaxy bias and the galaxy matter cross-correlation coefficient. We compute this bias, $C_{nl}$ directly 
		from simulations and then correct for it.
		
		The \mean{\eg}
		computed after applying the corrections, is within $\sim1\%$ of the 
		predicted value from the \lcdm\ model (compared to $5\%$ bias in the uncorrected $\mean{\eg}$). 
		A primary concern about these corrections is that they may depend on the details of HOD modeling and hence may not be very accurate.
		In appendix~\ref{append:hod_test}, we provide more tests using another set of 
		mocks with varying HODs and show that the corrections used for our main results ($z=0.25$ snapshot) 
		recover $\mean{\eg}$ to within $\lesssim2\%$. 
		Thus for our main results in the data, we will 
		use the corrections from the $z=0.25$ snapshot for both the full LOWZ and LOWZ-Z1 samples, the
        $z=0.40$ snapshot for the LOWZ-Z2 sample and $z=0.6$ snapshot for the CMASS sample.
		To account for the variations in these corrections, we also add $2\%$ systematic uncertainty in our error budget.
		        
	\subsection{$E_G$ measurement}\label{ssec:results_EG}
		\begin{figure}
        	 \centering
         	\includegraphics[width=\columnwidth]{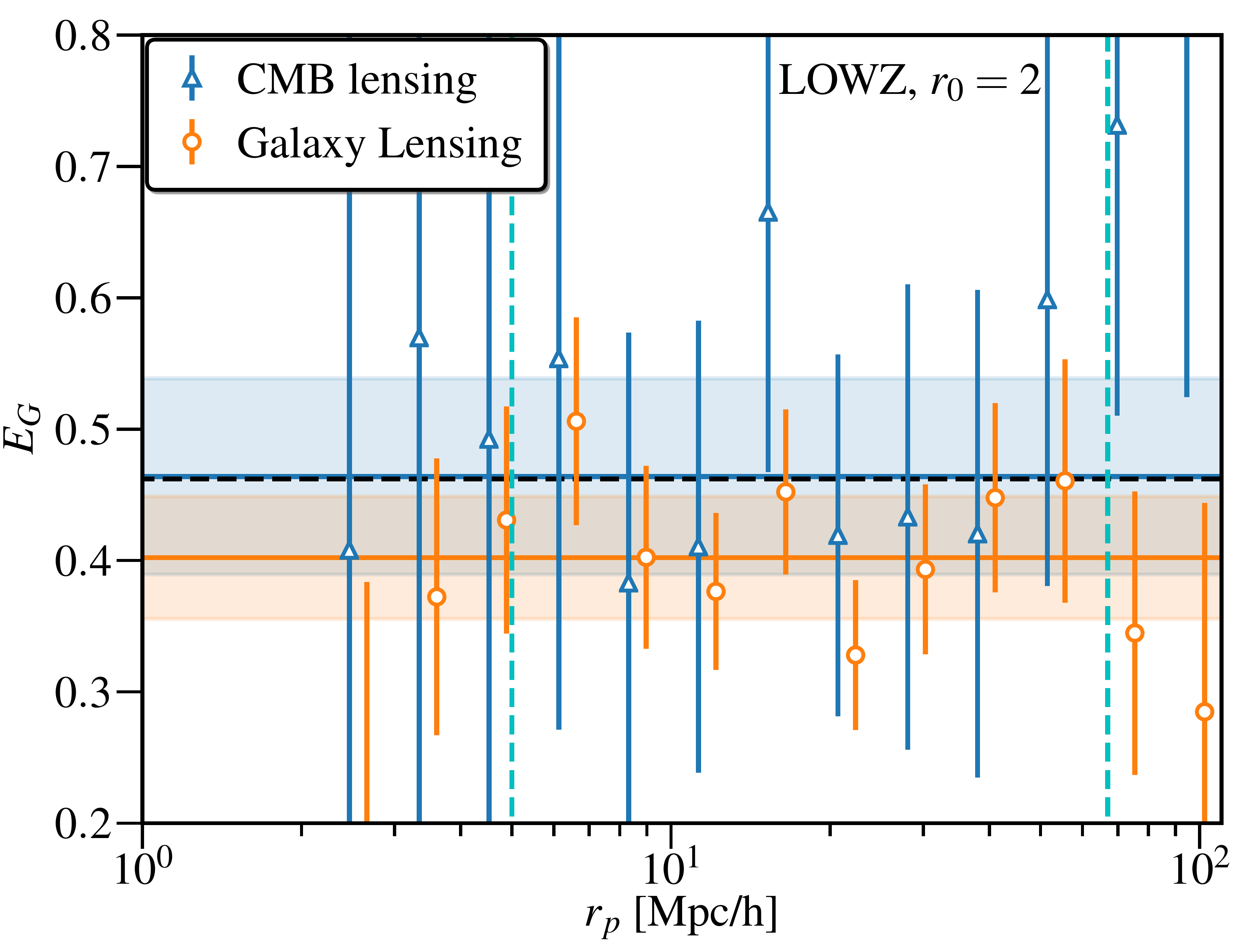}
	         \caption{\eg\ as a function of scale for LOWZ galaxies. The dashed black
	         line shows the Planck \lcdm\ prediction at $z=0.27$, while the solid 
	         lines 
	         show measured $\mean{\eg}$ over scales $5<r_p<70\mpch$. The bands show the
             $1\sigma$ errors on $\mean{\eg}$, averaged using the diagonal errors. Note that the galaxy lensing and CMB lensing are at different 
             effective redshift due to lensing weights, with galaxy lensing cross correlations at $z_\text{eff}=0.24$ and CMB lensing cross 
             correlations at $z_\text{eff}=0.3$. We apply correction $C_z$ to get final \eg\ at effective redshift of clustering, $z=0.27$.
	         }
    	     \label{fig:EG_lowz_cen}
        \end{figure}
        
         \begin{figure}
        		 \centering
	         	\includegraphics[width=\columnwidth]{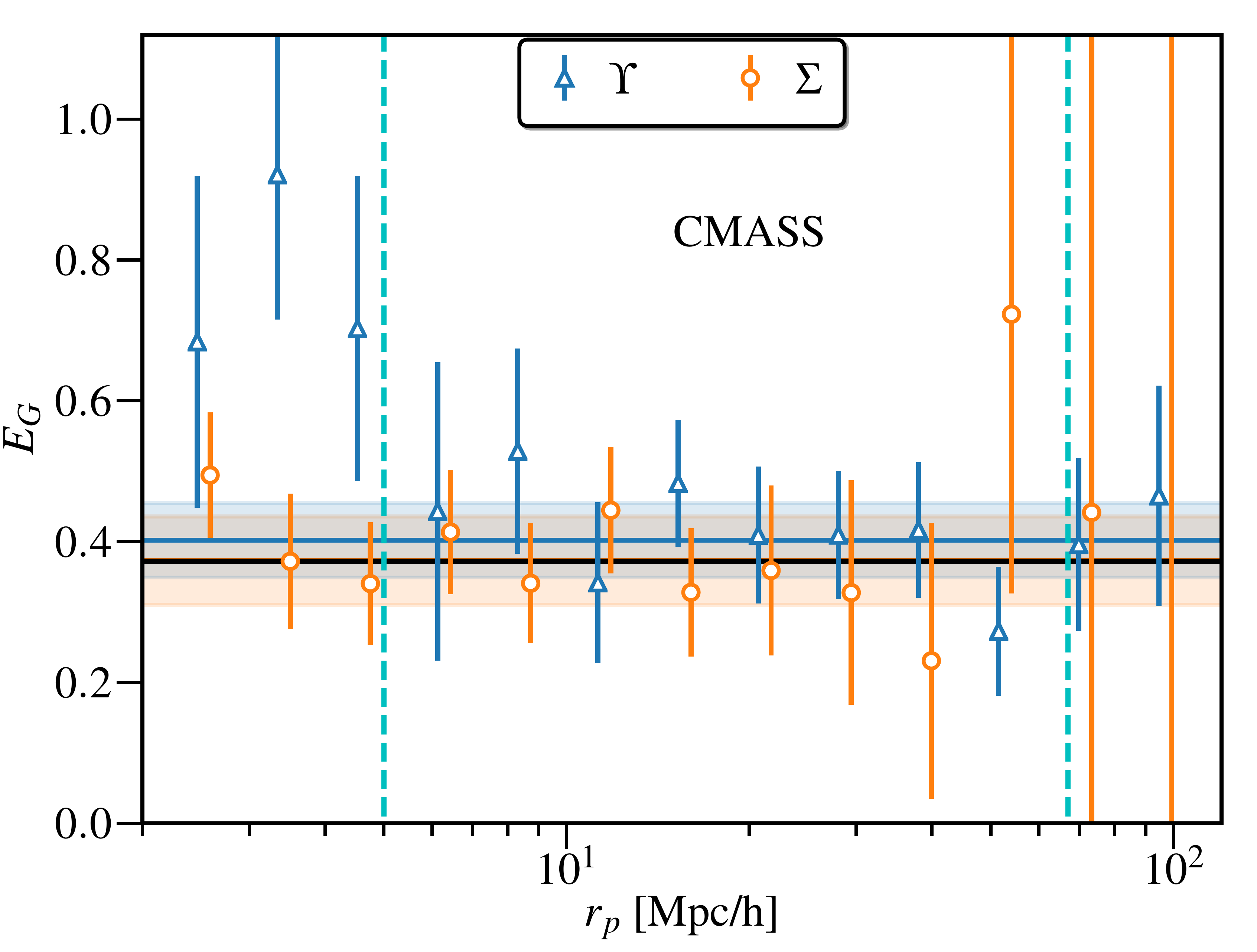}
		         \caption{ \eg\ measurement for the CMASS sample with CMB lensing. We show the
                   measurement using the $\Upsilon$ 
		         ($r_0=2\mpch$)
		         estimator as well as using $\Sigma$, for both CMB lensing and clustering. $\Sigma$
                 provides a 
		         better $S/N$ at small scales, while $\Upsilon$ has better $S/N$ at large scales due to reduced cosmic 
		         variance \citep{Baldauf2010}. The green band shows the measurement of the mean
                 \eg\ from \citealt{Pullen2016}, using the same 
		         datasets but in Fourier space, over the effective scale range $25$--$150\mpch$.  The dashed black
		         line shows the Planck \lcdm\ prediction.
		         }
	    	     \label{fig:EG_cmass}
    	    \end{figure}

        In Fig.~\ref{fig:EG_lowz_cen}, we show the $E_G$ measurement using the BOSS LOWZ sample along with both SDSS and
        CMB lensing maps. 
  		Using galaxy lensing, \mean{\eg} is 
        \referee{$\sim1.5\sigma$} low compared to Planck \lcdm\ predictions, 
        which is primarily driven by the low amplitude of 
        lensing measurements \referee{(which gets compensated partially by low $f$)}. 
        In the case of CMB 
        lensing, 
		\mean{\eg}
        is consistent with Planck \lcdm\ predictions, though the measurement is 
        noisier when compared to galaxy lensing. While the two measurements are statistically consistent, we note that galaxy-lensing 
        ($z_\text{eff}=0.24$) and CMB-lensing $z_\text{eff}=0.3$
        cross correlations are at different effective redshifts due to the impact of lensing weights and hence we apply different $C_z$ corrections to 
        get them at effective redshift of clustering sample. 
 
 		\eg\ measured from the field sample is consistent with the results from the full LOWZ sample
        (using the same corrections for 
		both). Since the field sample does not contain groups, the effects of non-linear bias and
        satellite contamination to the RSD measurement
		as well as to ratio of lensing and clustering (via its effects on $\rcc$)
		are expected to be smaller. 
        However, with the CiC cylinder size of $r_p\lesssim0.8\mpch$, we do not observe any 
		significant effects of removing groups at scales $r_p>5\mpch$, other than a reduced effective bias of the sample.
		CiC selection effects still bias the RSD measurements and hence we do not use the $f=0.74\pm0.05$
		measured from the field sample. 
        
        In Fig.~\ref{fig:EG_cmass}, we show the \eg\ measurement for the CMASS sample using the CMB
        lensing measurement with 
        two estimators, $\Sigma$ and $\Upsilon$, for clustering and lensing measurements. Both measurements are 
        consistent with Planck \lcdm\ predictions, though the mean \eg\ measured from $
        \Sigma$ is noisier compared to $\Upsilon$ due to different signal to noise as a function of scale.
        At small scales, $\Sigma$ gives better 
        signal-to-noise ratio ($S/N$) while at larger scales the $\Sigma$ measurement gets noisier. $\Upsilon$ on the 
        other hand 
        has lower $S/N$ at small scales since we are subtracting out some signal while at large scales it has better 
        $S/N$ because of a reduced contribution from cosmic variance (see discussion in section~\ref{ssec:formalism_EG}). 
        Note that corrections for CMASS sample are computed using the $z=0.6$ snapshot. 
        
        \begin{figure}
        		 \centering
	         	\includegraphics[width=\columnwidth]{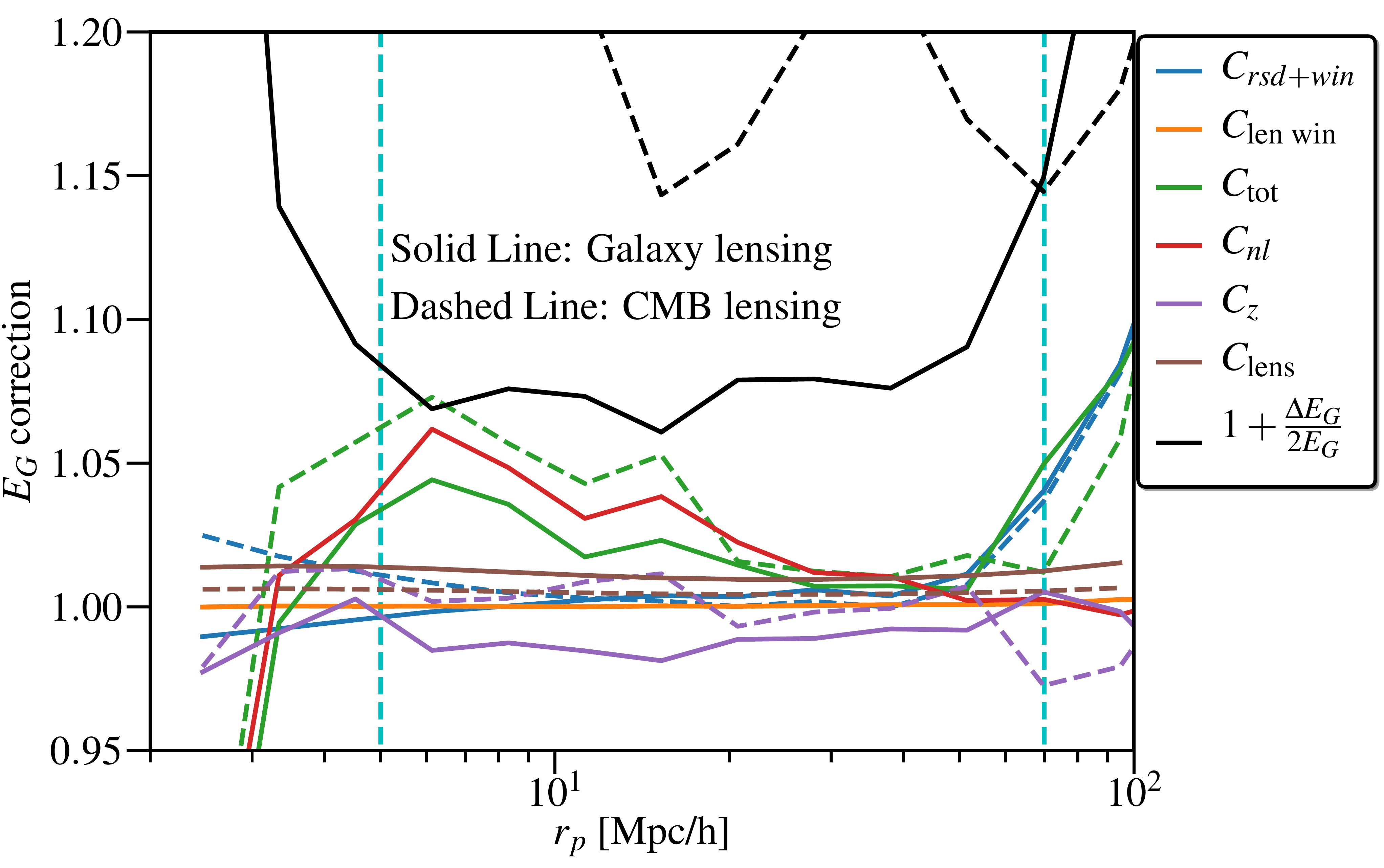}
		         \caption{Various correction terms applied to the \eg\ measurements in Fig.~\ref{fig:EG_lowz_cen} (LOWZ 
		       		  sample). The  black line also shows the relative uncertainty ({\em
                        rescaled} by a factor of two) in \eg\ as
                      a function of scale. The per-bin uncertainty for galaxy lensing is of order
                      15--20\% while for CMB lensing it is larger by a factor 
                      of 2 (30--40\%)
                        }
	    	     \label{fig:EG_corrections}
    	\end{figure}
		In Figure~\ref{fig:EG_corrections}, we show the magnitude of several corrections applied to
        the LOWZ \eg\ 
		measurements along with the relative (statistical) errors in the \eg\ measurement for comparison. 
		The $C_\text{rsd+win}$ correction is computed from a theoretical model (linear theory$+$Kaiser correction). 
		We also computed this correction using simulations (computing
		correlations with and without RSD) and find good agreement between theoretical estimates and simulations. 
		Simulations are noisier, hence we use the theoretical estimates in the final
        results. 

		All the scale-dependent corrections we applied are subdominant when compared to the
        statistical uncertainties in the \eg\ at that $r_p$, and the combined 
		effect of corrections is to change \eg\ by $\sim1-2\%$ ($\lesssim0.2\sigma$) for different samples. 
		\referee{$C_{nl}$ is the largest net correction on \mean{\eg} ($\sim 2-3\%$) as it affects
          small scales more with maximum values of order $5\%$, 
		though it reduces rapidly as scale increases. $C_{rsd+win}$ has a small impact as it is close to one on small 
		scales, though on larger scales (which are noisier and hence contribute less to \mean{\eg})
        it can be of order $15\%$. $C_{rsd+win}$ is large on large 
		scale because the correlation function 
		has small values at large scales and RSD effects, which are primarily additive, can lead to large relative changes. 
		The correction from the different 
		effective $z$ of clustering and lensing, $C_z$, is of order $1-1.5\%$ on small scales and
        $\sim 0.1\%$ on the largest scales. The correction due to 
		lensing magnification is $\sim 1-1.5\%$ and has only weak dependence with scale (see Figure~\ref{fig:lens_correction}) While 
		these corrections do not lead to a significant change in our results, their magnitude suggests that future work with much
        higher $S/N$ measurements 
		will have to model these corrections to higher precision.}
				
		\begin{figure}
        	 \centering
         	\includegraphics[width=\columnwidth]{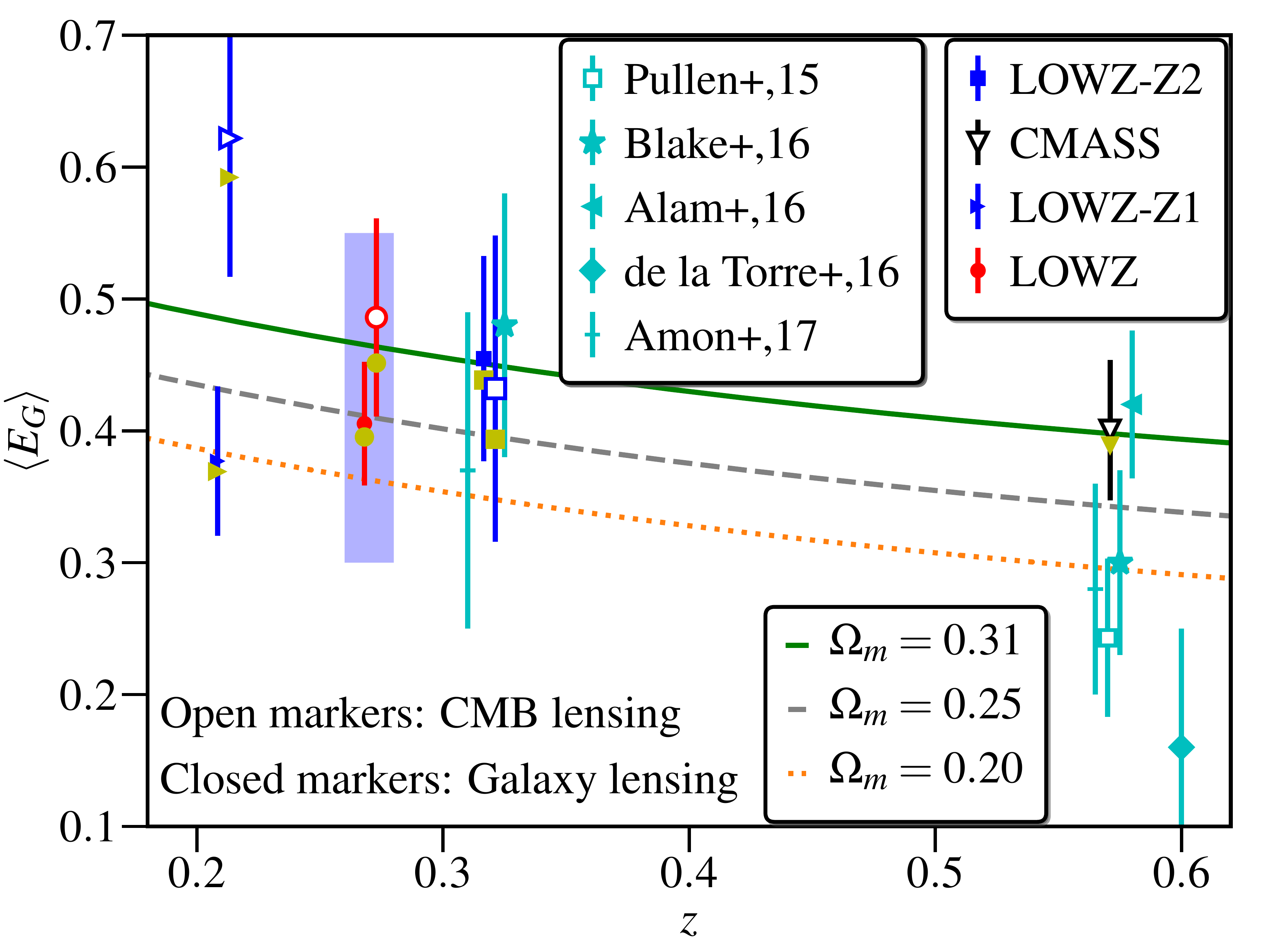}
	         \caption{\eg\ measurement between scales $5<r_p<70$ \mpch\ for
	         	different samples ({\em this work}), along with \lcdm\ predictions with different $
				\Omega_m$ values. Measurements in cyan are from other works, which may use a different range of scales. 
                  The measurements are consistent with the Planck \lcdm\ model ($\Omega_m=0.31$), though 
				given the size of the statistical uncertainties, it is not possible to rule out
                other models. The blue band around 
				the red points is drawn to indicate that the LOWZ measurement is correlated with the
                measurements from the Z1 and 
				Z2 samples. Solid yellow points show the uncorrected value of \mean{\eg} for the
                samples with measurements from this paper. Note that measurements from CMB and
                galaxy lensing are at different effective redshifts due to 
                the impact of lensing weights and hence separate $C_z$ corrections are applied to \eg\ using galaxy lensing and CMB lensing.
				}
    	     \label{fig:EG0_z}
        \end{figure}
		Finally, in Fig.~\ref{fig:EG0_z}, we show the redshift dependence of \mean{\eg} along with \lcdm\ predictions using
		different values of $\Omega_m$. The values of \mean{\eg} and $f$ are given in table~\ref{tab:params}. 
		In general our measurements for various samples are consistent with the Planck 
		\lcdm\ prediction at $2\sigma$ level or less. However the measurements are not sensitive enough to put meaningful 
		constraints on modified gravity parameters or to even rule out 
		lower values of $\Omega_m$, especially with galaxy-galaxy lensing measurements preferring somewhat lower 
		values. For the full LOWZ and CMASS samples, our measurements are consistent with those of 
		\cite{Blake2016} (using RCSlens and CFHTlens source galaxies), 
		who measured $\eg=0.48\pm0.10$ for LOWZ ($0.15<z<0.43$) and $\eg=0.30\pm0.07$ for the CMASS
		sample, with the measurement of \cite{Alam2016} (using CFHTlens source galaxies), $\eg=0.420\pm0.056$, for CMASS
        sample and also with \cite{Amon2018} (using KiDS), $\eg=0.37\pm0.12$ for LOWZ and $\eg=0.28\pm0.08$ for CMASS sample.

 		The measured amplitude of \eg\ is primarily affected by the (relative) lensing amplitude $A_l$ and the growth 
		rate, $f$, measurement from RSD. In the case of galaxy-galaxy lensing, the lensing amplitude is lower than 1, which drives the 
		measured \eg\ to be lower than the \lcdm\  predictions.
		The lensing measurement from galaxies is susceptible to systematic uncertainties in shear 
		estimation and photometric redshifts. As mentioned in Section~\ref{ssec:data_shapes}, 
		\cite{Reyes2012,Nakajima2012,Mandelbaum2013} did 
		extensive testing of the shape sample for these systematics and we use the calibration factors derived for the 
		shape sample derived in those papers, along with new corrections based on more recent
        simulations in HSC \citep{2017arXiv171000885M}. 
        We further test the accuracy of the calibration factors for photo-z in 
		appendix~\ref{append:clustering_pz} using a method with independent assumptions from the
        method used in \citet{Nakajima2012}.  There, the existence of a representative spectroscopic
        sample was assumed; here, we use the clustering redshift method, which assumes the existence
        of a non-representative spectroscopic sample that can be used to derive the ensemble
        redshift distribution by modeling the clustering signals.  In
        appendix~\ref{append:clustering_pz} we show that while the calibration factors from the
        clustering redshift method depend on
		the priors on the galaxy bias, they agree with the factors derived by \cite{Nakajima2012} to within $\sim 5\%$. 
		Using physically-motivated priors on the galaxy bias further improves the agreement. 

		There is also a possibility of contamination from intrinsic alignments (IA)
		of source galaxies. \cite{Blazek2012} estimated the IA contamination in the source sample used in this
		work and found no evidence for contamination when measuring the shear signal around LRG
        lenses,  with the conservative upper limit of the contamination being $5\%$.  
        We thus ignore the possible contamination from IA in this work. 
        
        We also note that the (relative) lensing amplitude is degenerate with galaxy bias which is fixed from clustering, 
        and lower apparent lensing amplitude does not necessarily mean 
        systematics in lensing alone. Any systematics in the projected clustering, \wgg\ , can also bias the \eg\ estimation and can show up as 
        low lensing amplitude if we are over estimating the clustering and hence the galaxy bias. Furthermore, systematics in clustering can also impact 
        the measurements of growth rate which in turn impact the \eg. While we do not find very significant deviations in the growth rate measurements, 
       	there are order 10\% variations in $f$ compared to the 
        predictions from Planck \lcdm\ cosmology and they drive \eg\ lower in case of $Z1$ sample and higher
        for $Z2$ and CMASS samples. 

		Another notable discrepancy is between our measurement for CMASS sample and that of \cite{Pullen2016}. 
		A significant part of the discrepancy in measurement of \cite{Pullen2016} (relative to Planck \lcdm\ prediction) is driven by large scales 
		($r_p\gtrsim80\mpch$), where the galaxy-CMB lensing amplitude was observed to be lower than
        expected in their measurements. Our
		results, on the other hand, are dominated by the measurements at small scales, with larger scales being noisier 
		and not contributing much to the mean \eg\ measurement. Even when considering the full
        scale-dependent \eg\ measurement, we do not observe any 
		significant deviations at large scales, though the uncertainties in the measurement at those scales are rather 
		large. 
		Using the scales $25<r_p<150\mpch$, we measure $\mean{\eg}=0.33\pm0.14$ using $\Sigma$ (statistically consistent with 
		\cite{Pullen2016} measurement of $0.24\pm0.06$ (stat) though errors are correlated between two studies) and 
		$\mean{\eg}=0.38\pm0.065$ using $\Upsilon$. We  note we use slightly different redshift range $0.45<z<0.7$ compared to \citep{Pullen2016} 
		($0.43<z<0.7$), but the volume in this range ($0.43<z<0.45$) is small, $<3\%$ of the sample, and is further down weighted due to rapidly 
		decreasing number density of galaxies as well as the CMB lensing kernel and hence should not contribute significantly to the 
		differences between the two 
		studies. Also, since we use (optimal)
        $\Sigma_\text{crit}^{-2}$ weighting in the galaxy-lensing cross-correlations, the noise properties of the measurements in the 
		two studies are somewhat different, though these differences should also be small.
		Another point to note is that \cite{Pullen2016}
        convert their scale-dependent 
		measurement to a \mean{\eg} using a  maximum likelihood (MLE) method that requires inversion
        of a noisy covariance 
		matrix. As a result the maximum likelihood point can be biased, which is accounted 
		for by widening the likelihood using Hartlap factor and hence the uncertainty in \mean{\eg}. For our results,
		we measure the \mean{\eg} in each jackknife region using only the diagonal covariance, which can also be suboptimal and introduce noise bias
		but we checked \referee{(by repeating the analysis with the full covariance)} that it does not 
		affect our results, as the off-diagonal elements of covariance matrix are expected to be subdominant. (This is because noise 
		is dominated by CMB 
		lensing reconstruction noise which acts as shot noise and hence the diagonal elements of 
		covariance should be dominant in both configuration and Fourier space.) 
		In essence, due to different treatment of 
		noise in covariances, the results in the two studies may not be very correlated. Using the jackknife fitting with diagonal 
		covariance on data from \cite{Pullen2016}, we find $\mean{\eg}=0.29\pm0.06$, 
		which further rises higher to $\mean{\eg}=0.32\pm0.06$ if we 
		ignore the lowest $\ell$ bin ( $\ell\lesssim70$) for which the errors are likely to be
        underestimated given the size of jackknife regions and also noise in error estimation (observed $S/N$ of this bin does not scale as the 
        expected $\sqrt{(2\ell+1)\Delta\ell}$ scaling relative to neighboring bins).  
		It is hard to further reconcile the measurements in the two works without doing additional tests that are outside the
		scope of this work.

\section{Conclusions}
	We have presented the measurements of \eg\ and its redshift dependence using BOSS galaxies and lensing measurements
	from SDSS galaxy lensing and Planck CMB lensing maps. Measurements from CMB lensing and galaxy lensing are 
	consistent within the noise for the LOWZ lens sample. With the higher redshift CMASS sample, the \eg\ measurement is of
	comparable significance to the galaxy lensing measurements for LOWZ sample. 
	This highlights the potential of CMB lensing to provide
	complimentary observations and strong consistency checks when combined with the galaxy lensing
    measurements \citep[see also][]{Singh2016,Schaan2016}.
	
	We also highlighted several theoretical uncertainties in computing \eg\ on nonlinear scales, and  computed corrections for 
	them. 
	Our results showed that after applying corrections in simulations we can recover \eg\ to
	about $2\%$ accuracy. With $\gtrsim10\%$ error in our measurement, these corrections are sufficient for this work. 
	However,
	in the future, several surveys will be able to measure \eg\ (or perform similar tests of GR) to sub-percent accuracy
	\citep{Leonard2015,Pourtsidou2016}. 
	For these surveys, it will be important to compute the corrections for theoretical uncertainties
    to even higher accuracy, 
	using better theoretical modeling with analytical models and/or simulations.
	
	While our \eg\ measurements are largely consistent with the predictions from Planck \lcdm\ predictions, there are 
	some deviations at $\lesssim2\sigma$ level, 
	especially when using galaxy lensing measurements. Though statistically not very significant in \eg, 
	these deviations are
	primarily driven by the low amplitude of lensing measurements, which can possibly be due to systematics in either 
	clustering and/or lensing measurements, since clustering bias and lensing amplitude are
    degenerate. The lensing 
	amplitude is also degenerate with the $\sim\sigma_8\Omega_m^{1/2}$ value, and low amplitude could also mean that data 
	prefers lower
	$\Omega_m$ or $\sigma_8$ (or both) values compared to the Planck \lcdm\ model assumed.
	We also see similar redshift-dependent deviations in the growth rate measurements (and measurements with CMB
	lensing), though the uncertainties are too large to make a definitive statement. 
	Since the growth rate is also degenerate with galaxy bias, this redshift-dependent deviation does suggest that it is 
	possible that the problem (at least partly) is from the some residual sample selection effects which affect the
    galaxy clustering measurements.  
    
    Low amplitude lensing measurements (compared to predictions from Planck \lcdm\ model) have
    also been observed by other lensing studies (see for example \cite{Hildebrandt2016,Leauthaud2017,Joudaki2017,DES2017} 
     and for results consistent with Planck cosmology, \citealt{Uitert2017}). 
    \cite{Leauthaud2017} 
    performed tests for effects of several systematics and the physics beyond \lcdm\ model in galaxy-galaxy lensing 
    (eg. effects of baryonic physics, neutrinos, 
    modified gravity, assembly bias, sample selection) and showed that such effects 
    can be significant, especially with several systematics and physical effects being combined together. 
    \cite{Joudaki2017} performed a similar analysis for cosmic shear measurements and also showed that 
    marginalizing over some models for systematics can relieve some of the tensions between Planck \lcdm\ model and 
    lensing measurements. These studies, along with our tests, suggest that more work is required to 
    study and model the impact of systematics in both lensing and clustering measurements.
	  
	In the near future, data from the eBOSS, DES, KiDS and HSC surveys will help to extend the growth rate and \eg\ 
	measurements to 
	higher redshifts with measurement uncertainties likely to be around $\lesssim5\%$ level. 
	With the advent of LSST, DESI, WFIRST, SKA and CMB Stage-IV surveys, the statistical uncertainties on \eg\ will 
	decrease considerably, providing percent level or better measurements. However, to make \eg\ a strong test of 
	gravity and \lcdm\ it is imperative to 
	improve the modeling to mitigate observational systematics as well as theoretical uncertainties to much higher 
	precision than was done in this work.

\section*{Acknowledgements}
    We thank Martin White and Beth Reid for providing us with a halo catalog from simulations, Anthony Pullen for providing his \eg\ measurements
    and 
	the SDSS-I/II/III and Planck collaboration for their efforts in providing the datasets used in this 
    work. We also thank Danielle Leonard for useful discussions and comments on the draft version of
    the paper. \referee{We thank the anonymous referee for  
    providing comments that helped in improving the quality of the paper.}
    
    RM acknowledges the support of the Department of Energy Early Career Award program and grant DE-SC0010118.
    SS acknowledges support from John Peoples Jr.\ Presidential Fellowship from Carnegie Mellon University. 
    SA is supported by the European Research Council through the COSFORM Research Grant (\#670193).
    
    Some of the results in this paper have been derived using the HEALPix package \citep{Gorski2005}.
    
    Funding for SDSS-III has been provided by the Alfred P. Sloan Foundation, the Participating Institutions, the 
    National Science Foundation, and the U.S. Department of Energy Office of Science. 
    The SDSS-III web site is http://www.SDSS3.org/.
    
    SDSS-III is managed by the Astrophysical Research Consortium for the Participating Institutions of the SDSS-III 
    Collaboration including the University of Arizona, the Brazilian Participation Group, Brookhaven National 
    Laboratory, Carnegie Mellon University, University of Florida, the French Participation Group, the German 
    Participation Group, Harvard University, the Instituto de Astrofisica de Canarias, the Michigan State/Notre Dame/
    JINA Participation Group, Johns Hopkins University, Lawrence Berkeley National Laboratory, Max Planck Institute for 
    Astrophysics, Max Planck Institute for Extraterrestrial Physics, New Mexico State University, New York University, 
    Ohio State University, Pennsylvania State University, University of Portsmouth, Princeton University, the Spanish 
    Participation Group, University of Tokyo, University of Utah, Vanderbilt University, University of Virginia, 
    University of Washington, and Yale University.

\bibliographystyle{mnras}
        \bibliography{papers,sukhdeep,shadab}

\appendix
\onecolumn
\section{Lensing window function}\label{append:lensing_window}

We define the window function as the weight assigned in the galaxy mass correlation function (Eq.~\eqref{eq:losintgm}) at a given 
line-of-sight separation 
from lens galaxy. 
For a given lens-source pair, with a lens at $z_l$ and source at $z_s$ with photometric
redshift $z_p$, the (unnormalized) window function is given by
	\begin{equation}
		w_\text{win}(\Pi|z_l,z_s,z_p)=\Sigma_\text{crit}^{-2}(z_l,z_p)\int_0^{z_s} dz\, 
		\Sigma_\text{crit}^{-1}(z,z_s)\Sigma_\text{crit}(z_l,z_p) \delta_D\left(z-z_l-\frac{\Pi H(z_l)}{c}\right)
	\end{equation}
	where $\Pi$ is the line-of-sight separation from the lens. $\delta_D$ is the dirac delta
    function and enforces the correct relationship between $\Pi$ and $z$.
	$\Sigma_\text{crit}^{-2}(z_l,z_s)$ factor arises from the weights we used in $\Delta\Sigma$ estimator. 
	$\Sigma_\text{crit}^{-1}(z,z_s)$ is the true critical surface density at $z$ that weighs the contribution of matter 
	fluctuation to shear, while 
	$\Sigma_\text{crit}(z_l,z_p)$ is the critical surface density that we use to convert the shear back to matter density 
	when measuring the signal. Since the two $\Sigma_\text{crit}$ factors are not the same, contributions from matter fluctuations
	at different redshifts are weighted differently in the correlation function leading to a non-trivial window function.

	Integrating over the lens and source samples we get
	\begin{equation}
		w_\text{win}(\Pi)=\frac{1}{\mathcal N}\int dz_l P(z_l) f_k(\chi_l)^{-2}\int_{z_l}^\infty dz_p P(z_p)\Sigma_\text{crit}^{-2}(z_l,z_p)
		\int_0^2dz_sP(z_s|z_p) \int_0^{z_s} dz
		\Sigma_\text{crit}^{-1}(z,z_s)\Sigma_\text{crit}(z_l,z_p) \delta_D\left(z-z_l-\frac{\Pi H(z_l)}{c}\right)
	\end{equation}
	where the normalization factor $\mathcal N$ is defined such that $w(\Pi=0)=1$. Also note that
    the effects of bias and scatter in 
	photometric redshifts are included in the calibration biases derived in \cite{Nakajima2012} and hence we do not 
	include them in our calculation. We set $P(z_s|z_p)=\delta_D(z_s-z_p)$ to get the final expression  
	\begin{equation}
		w_\text{win}(\Pi)=\frac{1}{\mathcal N}\int dz_l P(z_l) f_k(\chi_l)^{-2}\int_{z_l}^\infty dz_p P(z_p)\Sigma_\text{crit}^{-2}(z_l,z_p)
		\int_0^{z_p} dz
		\Sigma_\text{crit}^{-1}(z,z_p)\Sigma_\text{crit}(z_l,z_p) \delta_D\left(z-z_l-\frac{\Pi H(z_l)}{c}\right)
	\end{equation}

\section{Estimating lensing magnification bias to $E_G$}\label{append:lensing_EG}
	As shown by \cite{Dizgah2016}, the effects of lensing magnification modify the clustering and lensing observables and hence 
	can bias the \eg\ measurements. In this section we derive impact of our failure to model magnification bias on our real space 
	clustering, lensing and \eg\ observables.

	\subsection{Bias in Clustering}
		Due to the effect of lensing by the foreground structure, the apparent galaxy over-density is modified from the 
		true over-density. This modification arises from two effects, the volume perturbations due to magnification, and 
		the modification to the galaxy selection function \citep{Yoo2009,Bernstein2009}. Most previous studies have derived the magnification bias 
		to clustering in Fourier space. In this section we derive the expression for magnification bias in projected clustering in real space with 
		a limited line-of-sight window function.
		
		Following the notation of \cite{Hildebrandt2009}, to estimate the impact of lensing contamination on the galaxy clustering auto-correlation, 
		we begin by assuming that the true number density of galaxies, 
		$n_0$, for a given flux limit, $f$, follows a simple power-law relation
		\begin{equation}
			n_0(>f)=Af^{-\alpha},
		\end{equation}
		 where the slope $\alpha$ is measured as 
		\begin{align}
			\alpha=-\frac{ d\log{(n_0(>f))}}{d \log{(f)}}=2.5\frac{d\log{(n_0(<m))}}{dm}.
		\end{align}
		where in the second equality we converted flux to the magnitude, $m$.
		
		The magnification $\mu$ modifies the observed volume and the galaxy flux by $\frac{\widehat{V}}{V}=\frac{\widehat{f}_g}{f_g}\sim \mu$. Thus
		the observed number density of galaxies, $\widehat n$, changes due to the magnification $\mu$ as
		\begin{align}
			\widehat n(>f)=&\frac{V}{\widehat V}n_0(>\frac{f}{\mu})=\frac{1}{\mu}n_0(>\frac{f}{\mu})\nonumber\\
				 =&\frac{1}{\mu}\mu^\alpha n_0(>{f})\nonumber\\
			\widehat n(>f)\approx&(1+2(\alpha-1)\kappa) n_0(>{f})
		\end{align}
		where we used the relation between magnification and convergence, $\kappa$, $\mu\approx1/(1-\kappa)^2$. 

		The observed galaxy over-density, $\widehat\delta_g$, is then related to the true 
		over-density, $\delta_g$, to first order as
		\begin{equation}
			\widehat\delta_g\approx\frac{(1+\kappa') n_g}{\overline{n}_g}-1\approx\delta_g+\kappa',
		\end{equation}
		where $\kappa'=2(\alpha-1)\kappa$.
	
		The measured correlation function is then
		\begin{align}
			\widehat{\xi}_{gg}(r_p,\Pi)&=\xigg(r_p,\Pi)+
			\int dz\, W(z)\mean{\kappa'(z)\kappa'{(z+\Delta z_\Pi)}}
			\left(\frac{r_p}{f_k(\chi_z)}\right)\nonumber\\
			&+\int dz\, W(z)\mean{\delta_g(z)\kappa'(z+\Delta z_\Pi)}(r_p)+\int dz\, W(z)\mean{\delta_g(z-\Delta z_\Pi)\kappa'(z)}(r_p)
			\label{eq:xi_magnification}
		\end{align}
		where the integrals are over the redshift range of the galaxy sample. $\kappa'(z)$ refers to
        the (magnification) convergence for a source at $z$, 
		$\kappa'(z)\kappa'{(z+\Delta z_\Pi)}(\frac{r_p}{f_k(\chi_z)})$
		is the convergence correlation
		function at angular separation $\theta=r_p/f_k(\chi_z)$ for sources at $z$ and $z+\Delta
        z_\Pi$ and 
		$\delta_g\kappa'$ is the cross correlation between the galaxy over-density field and
		 convergence. The projected correlation function is then
		\begin{align}
			\widehat{w}_{gg}(r_p)&=\wgg(r_p)+\int_{-\Pi_\text{max}}^{\Pi_\text{max}}d\Pi
			\int dz\,  W(z)\mean{\kappa'(z)\kappa'{(z+\Delta z_\Pi)}}
			\left(\frac{r_p}{f_k(\chi_z)}\right)\nonumber \\
			&+\int_{-\Pi_\text{max}}^{\Pi_\text{max}}d\Pi\int dz\,  W(z)
			\mean{\delta_g(z)\kappa'(z+\Delta z_\Pi)}(r_p)\\
			&=\wgg(r_p)+[2(\alpha-1)]^2
			\int_{-\Pi_\text{max}}^{\Pi_\text{max}}d\Pi\int dz\,  W(z)\int_0^{z+\Delta z} d\chi_m
			\frac{\overline\rho_m^2 w_{mm}\left(r_p\frac{f_k(\chi_{z_m})}{f_k(\chi_z)},z_m\right)}{\Sigma_\text{crit}(z_m,z)
			\Sigma_\text{crit}(z_m,z+\Delta z)}\nonumber \\
			&+2(\alpha-1)\int_{-\Pi_\text{max}}^{\Pi_\text{max}}d\Pi\int dz\, W(z){\overline{\rho}_m w_{gm}(r_p,z)}
			\left(\frac{1}{\Sigma_\text{crit}(z,z+\Delta z)}+\frac{1}{\Sigma_\text{crit}(z-\Delta z,z)}\right)
		\end{align} 
		where $\Delta z_\Pi=\frac{\Pi H(z)}{c}$ and $p_{g}(z)$ is the galaxy redshift distribution function. 
			Under the assumption that $f(\chi_z)\gg \Pi_{max}$, the last term can be approximated as
		\begin{align}
			\int_{-\Pi_\text{max}}^{\Pi_\text{max}}d\Pi\int dz\, W(z){\overline{\rho}_m w_{gm}(r_p,z)}
			&\left(\frac{1}{\Sigma_\text{crit}(z,z+\Delta z)}+\frac{1}{\Sigma_\text{crit}(z-\Delta z,z)}\right)\nonumber \\
			&\approx2\int dz\, W(z)\overline{\rho}_m w_{gm}(r_p,z)
			\frac{4\pi G (1+z)}{c^2}
			\int_{0}^{\Pi_\text{max}}d\Pi \frac{f_k(\chi_z)\Pi}{f_k(\chi_z)+\Pi}\\
			&=2\int dz\, W(z)\overline{\rho}_m w_{gm}(r_p,z)
			\frac{4\pi G (1+z)}{c^2}
			\left[f_k(\chi_z)\Pi_\text{max}-f_k(\chi_z)^2\ln\left(\frac{f_k(\chi_z)+\Pi_\text{max}}{f_k(\chi_z)}\right)\right]\nonumber\\
			&\approx2\frac{3H_0^2\Omega_m}{2c^2}\frac{\Pi_{max}^2}{2}\int dz\, W(z)(1+z)w_{gm}(r_p,z)\\
			&\approx5\times10^{-4}\int dz\, W(z)(1+z)w_{gm}(r_p,z)\nonumber\\
			&\approx5\times10^{-4}(1+\overline{z})w_{gm}(r_p)
		\end{align}
			Note that our estimation of this term is $\sim2$ orders of magnitude lower than the estimation by \cite{Dizgah2016}. This is likely 
			driven by the different choice of $\Pi_\text{max}$, as \cite{Dizgah2016} use $C_\ell$
			where large effective line-of-sight integration length 
			can give
            $\sim2$ orders of magnitude difference in the contamination due to unmodeled
            magnification effects.

\begin{figure*}
        	\centering
			\begin{subfigure}{.45\columnwidth}		
         		\includegraphics[width=\columnwidth]{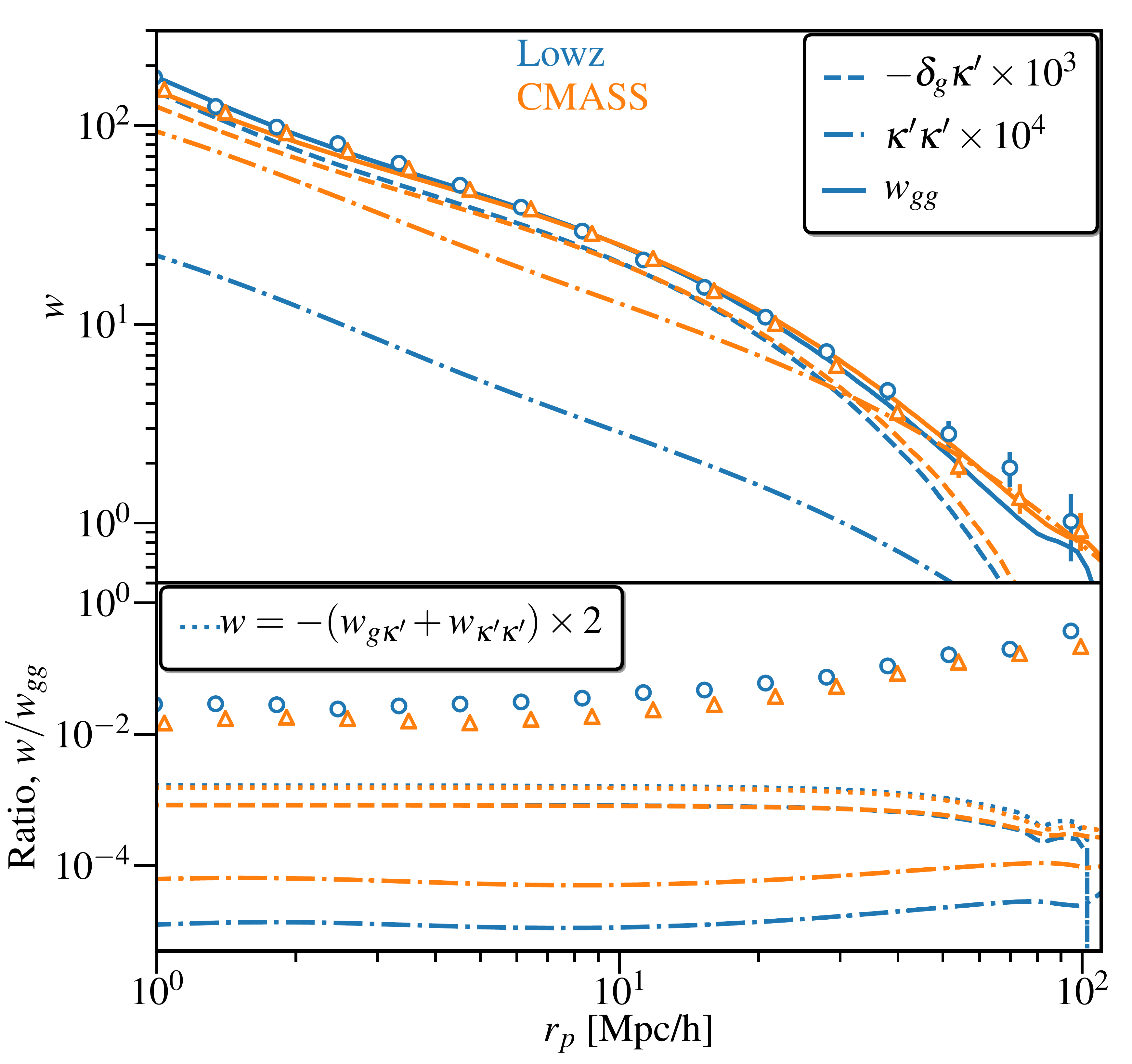}
		         \caption{Clustering	}
	    	     \label{fig:lens_correction_wgg}
		     \end{subfigure}
	    	 \begin{subfigure}{.45\columnwidth}		
         		\includegraphics[width=\columnwidth]{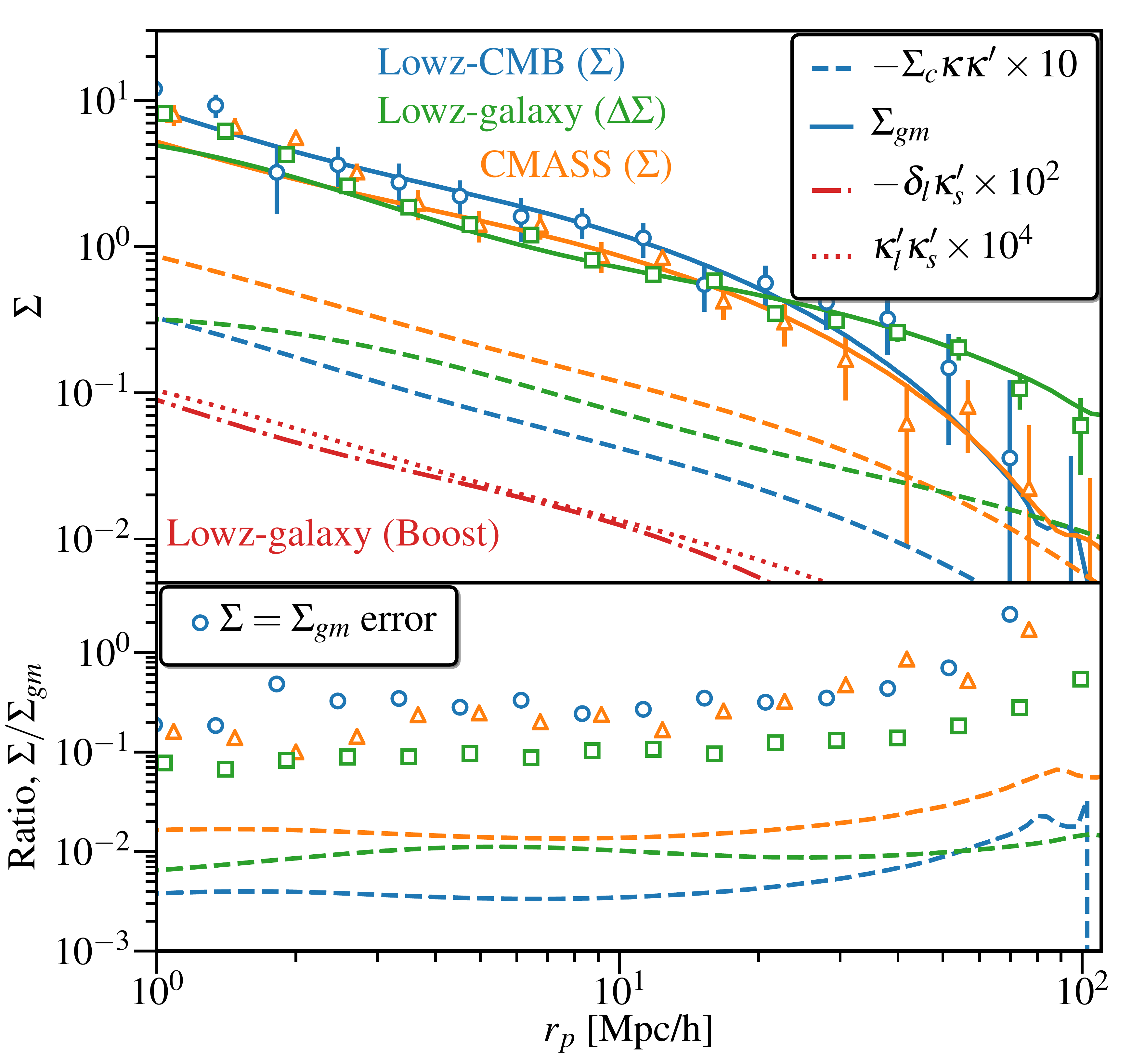}
	        	 \caption{Lensing}
	    	     \label{fig:lens_correction_wl}
		     \end{subfigure}
	     	\caption{a) Lensing contamination to the clustering measurements. The upper panel shows the correlation function, 
	     		\wgg, and the correlations due to lensing ${g\kappa'}$ and
                ${\kappa'\kappa'}$. The lower panel shows the 
				ratio of predicted contamination with the best fit model. The dotted lines in the lower panel are the total 
				contamination from two terms, rescaled by a factor of two for clarity.
				Also, the open markers show the noise-to-signal
				ratio in the measurements. 
                 Note that in the upper panel (but not in lower panel, unless 
                 stated in 
				lower panel legend), we multiplied 
				lensing predictions with a constant factor to make them same order of magnitude as the clustering 
				signal.	 
	     	b) Same as a), but showing the lensing contamination to $\Sigma$ and  $\Delta\Sigma$ measurements.
	     }
	     \label{fig:lens_correction}
        \end{figure*}

		In Fig.~\ref{fig:lens_correction_wgg} we show the estimated contamination from lensing to the clustering
		measurement for the case of both LOWZ and CMASS samples. 
		We estimated $\alpha=0.4$ for LOWZ and $\alpha=0.5$ for CMASS using the slope of luminosity function on the fainter end.
		The dominant contribution is from ${\delta_g\kappa'}$, which biases
		the clustering low by a factor of $\sim10^{-3}$, much less than the statistical errors. 

		\subsubsection{Bias in redshift-space multipoles}
\begin{figure*}
            	\centering
    			\begin{subfigure}{.45\columnwidth}		
             		\includegraphics[width=\columnwidth]{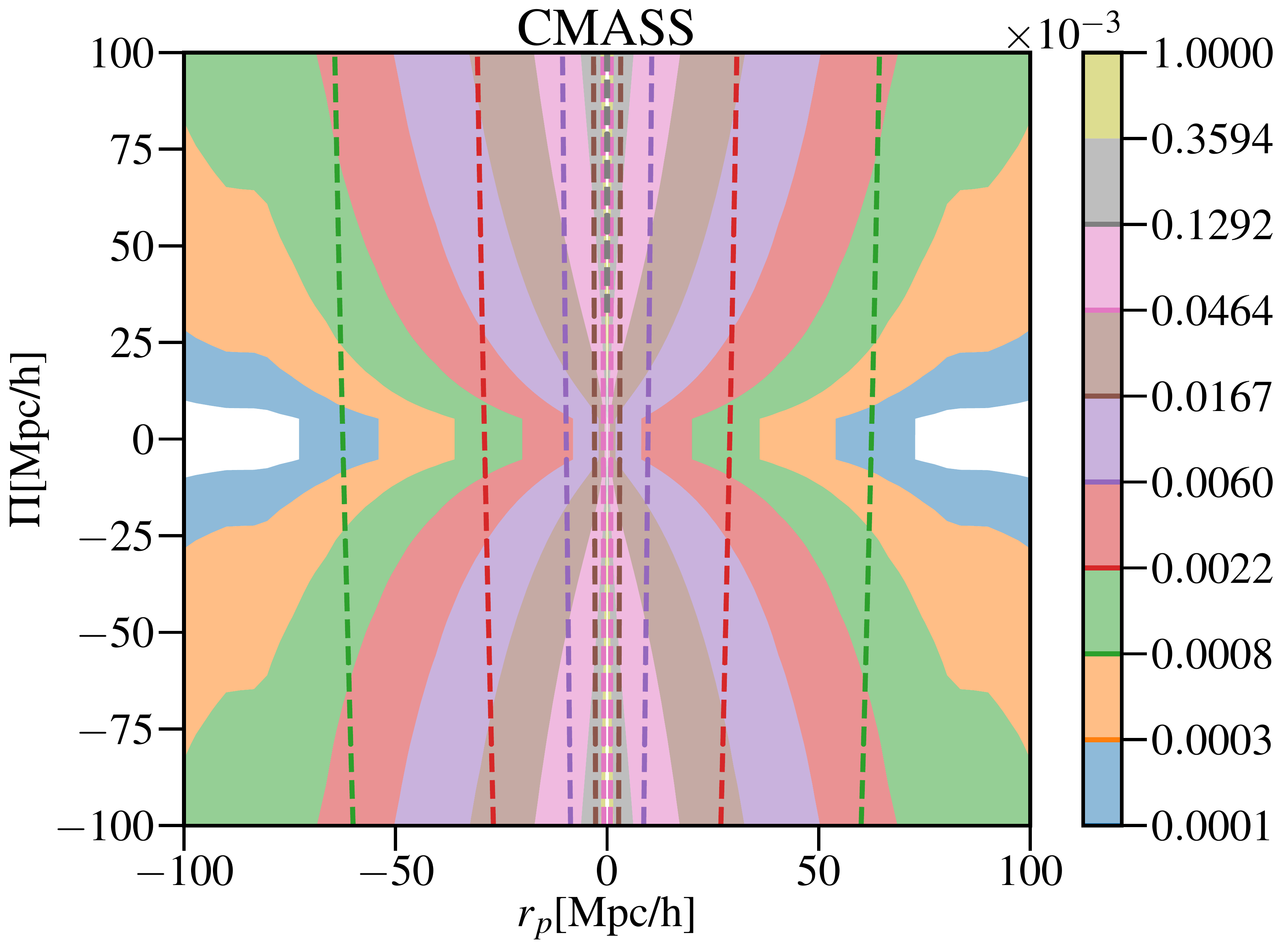}
    		         \caption{}
    	    	     \label{fig:RSD_lens_2D}
    		     \end{subfigure}
    	    	 \begin{subfigure}{.45\columnwidth}		
             		\includegraphics[width=\columnwidth]{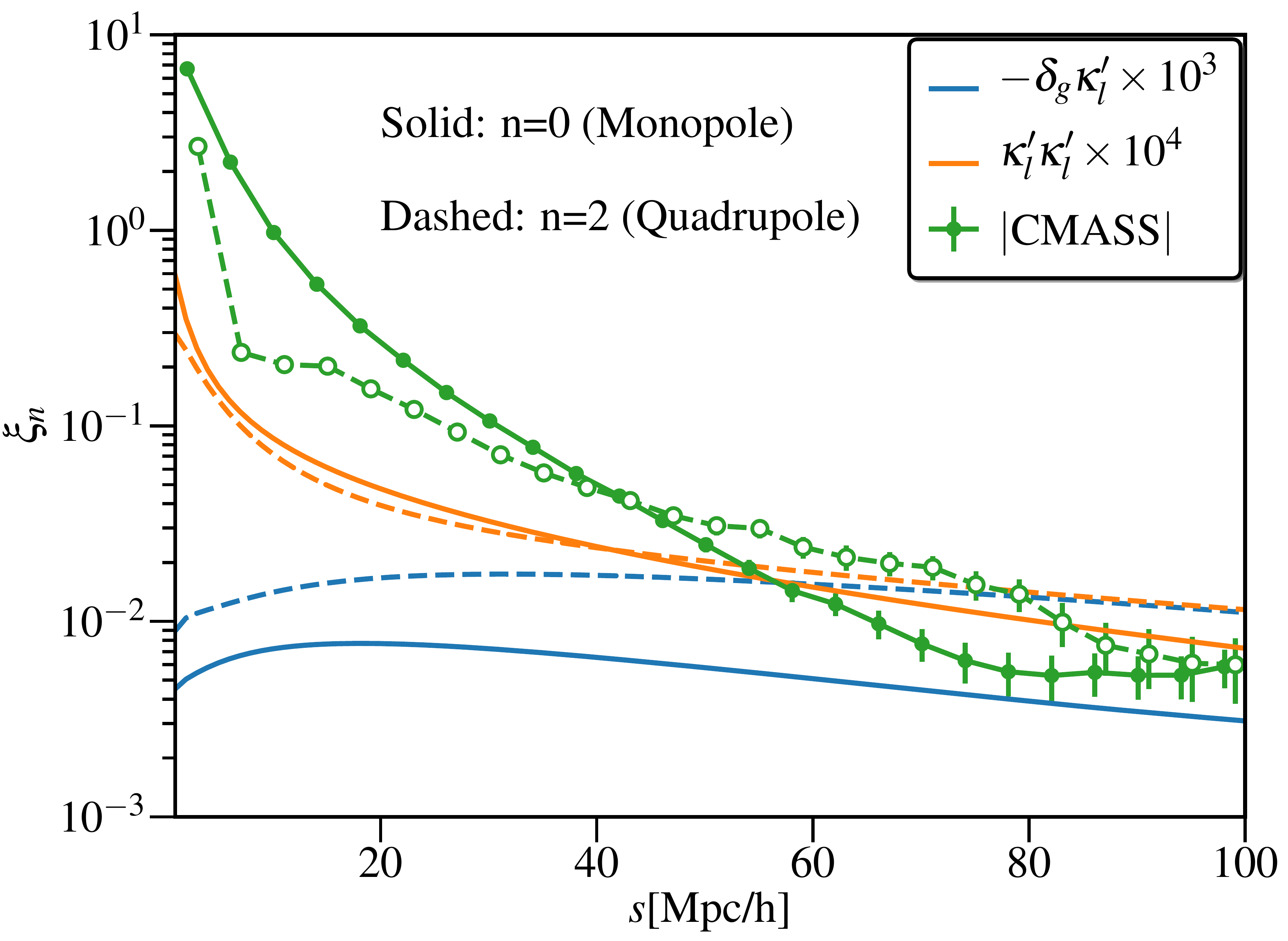}
    	        	 \caption{}
    	    	     \label{fig:RSD_lens}
    		     \end{subfigure}
    	    	\label{fig:rsd_lens_correction}
    	     	\caption{a) Contamination from magnification to clustering  as a function of
                  projected and line-of-sight separations. Solid contours are the $\delta_g\kappa'$
                  term with a negative sign and dashed 
					lines are the $\kappa'\kappa'$ term.
    	     	b)  Contamination from magnification to clustering multipoles.
    	     }
            \end{figure*}

			In Fig.~\ref{fig:RSD_lens_2D} we show the magnification terms from
            Eq.~\eqref{eq:xi_magnification} as a function of projected and line-of-sight 
			separations. Due to the dependence of the lensing kernel on distances, these terms show anisotropic structure. The galaxy-magnification 
			cross correlation increases at a given projected scale as the line-of-sight separation increases because
			the lensing kernel 
			becomes more sensitive to the galaxies (i.e.\ the lensing kernel increases at the position of galaxies). 
			The magnification auto-correlation increases with the mean redshift of the pair of galaxies, 
			which increases with $\Pi$. As a result both of these terms also contribute to the multipoles of the correlation function, though as 
			discussed in the previous section, these contributions are order $0.1\%$ or smaller when compared to the clustering signals. 
			The monopole and quadrupole terms 
			are shown in Fig.~\ref{fig:RSD_lens}. In addition, these terms also have dipole
            contributions with the magnification autocorrelation being 
			larger though it is smaller than the monopole by an order of magnitude.

	\subsection{Bias in lensing}\label{append_ssec:lens_bias}

		We can model the magnification bias in galaxy-lensing cross correlations in a similar way to clustering, 
		as the magnification bias is changing the 
		observed density contrast of the lens and source samples.  
          For the case of galaxy-galaxy lensing
		\begin{equation}
			\widehat{\Delta\Sigma}(r_p)=\langle\widehat b_F(r_p)\Sigma_\text{crit}\widehat\delta_g\widehat\gamma_t\rangle(r_p)
		\end{equation}
		where $b_F(r_p)=\frac{\sum_{ls}{w_{ls}(r_p)}}{\sum_{Rs}{w_{Rs}(r_p)}}\approx1$ 
		is the boost factor and is affected by the lensing magnification of both lens 
		and source samples. We will derive the contamination in boost factor, $\delta b_F$ ($\widehat b_F=b_F+\delta b_F$), later in the section.
		 
		Since the shear is 
		measured at the position of source galaxies, $\widehat{\theta}$, it can be written as
		\begin{equation}
			\widehat{\gamma}(\widehat\theta)=(1+\delta_{g,{s}}+\kappa'_{s})\gamma (\widehat\theta)
		\end{equation}
		The cross correlation with the lens sample is 
		\begin{equation}
			\mean{\widehat{\gamma}\widehat{\delta}_{g,l}}=\mean{{\gamma}{\delta}_{g,l}}
							+\mean{\gamma\kappa'_l}+\mean{\gamma\kappa'_\text{s}\kappa'_l}
							+\mean{\gamma\kappa'_{s}\delta_{g,l}}
		\end{equation}
		In our calculations we will ignore the terms at third order. The measured galaxy lensing signal is 
		\begin{equation}
			\widehat{\Delta\Sigma}(r_p)
			=\Delta\Sigma(r_p)+\langle \widehat{b}_F\Sigma_\text{crit}\gamma_t\kappa_l'\rangle(r_p)+\delta b_F(r_p)\Delta\Sigma(r_p)
		\end{equation}
		
		For the case of galaxy-CMB lensing, the magnification contamination is 
		\begin{equation}
			\widehat{\Sigma}(r_p)=\langle\Sigma_\text{crit}\widehat\delta_g\kappa_\text{cmb}\rangle(r_p)=\Sigma(r_p)+
			\langle\Sigma_\text{crit}\kappa_\text{cmb}\kappa_l'\rangle(r_p)
		\end{equation}

		In Fig.~\ref{fig:lens_correction_wl}, we show the estimated contamination in lensing measurements using full calculations 
		incorporating lens-source redshift distributions. Magnification contamination 
		introduces a bias of order $\sim1\%$ in $\Delta\Sigma$. Note that this contamination can increase for high lens redshifts 
		as the $\kappa\kappa$ or $\gamma\kappa$ term becomes larger. 
	
	\subsubsection{Bias in boost factor}
	
		Source galaxies that are physically associated with lens galaxies do not contribute to the lensing shear but do lower the normalization of 
		the 
		lensing measurement around lens galaxies. As described in Section~\ref{sssec:estimator_galaxy_galaxy_lensing}, to account for this effect
		 the final lensing measurement is multiplied by the boost factor, $b_F$, defined as
		\begin{equation}
			b_F(r_p)=\frac{\sum_{ls}w_{ls}(r_p)}{\sum_{Rs}w_{Rs}(r_p)}=\frac{1}{N}\int dz_lp(z_l)f_k(\chi_l)^{-2}\int dz_p p(z_p)
			\Sigma_{c}^{-2}(z_l,z_p)\int dz_sp(z_s|z_p)(1+\mean{\delta_l\delta_s}(r_p))
		\end {equation}
		where $\Sigma_{c}^{-2}(z_l,z_p)$ arises due to lens-source pair weighting,
        $w_{ls}$. $f_k(\chi_l)^{-2}$ acts as an effective lens weight as
		the number of source galaxies contributing to a fixed comoving bin decreases with increasing lens redshift (in other words, the 
		comoving projected number density of source galaxies in the lens plane decreases with lens redshift).
		
		However, in the presence of magnification, the number density of source galaxies around lenses changes, modifying the boost factor as
		\begin{align}
			\widehat{b}_F(r_p)=&\frac{1}{N}\int dz_lp(z_l)f_k(\chi_l)^{-2}\int dz_p p(z_p)\Sigma_{c}^{-2}(z_l,z_p)\int dz_sp(z_s|z_p)\left[1+
			\mean{(\delta_l+\kappa'_l)(\delta_s+\kappa'_s)}(r_p)\right]\nonumber\\
			=&b_F(r_p)+\frac{1}{N}\int dz_lp(z_l)f_k(\chi_l)^{-2}\int dz_p p(z_p)\Sigma_{c}^{-2}(z_l,z_p)\int dz_sp(z_s|z_p)
				\left[\mean{\kappa'_l\delta_s+
				\delta_l\kappa'_s+\kappa'_l\kappa'_s}(r_p)\right]\nonumber\\
			\approx &b_F(r_p)+\frac{1}{N}\int dz_lp(z_l)f_k(\chi_l)^{-2}\int dz_p p(z_p)\Sigma_{c}^{-2}(z_l,z_p)\int dz_sp(z_s|z_p)
				\left[\mean{\delta_l\kappa'_s+\kappa'_l\kappa'_s}(r_p)\right]
		\end {align}
		Here we assume that source galaxies are always behind the lens and hence ignore the $\kappa'_l\delta_s$ term in the last
		equality. Since the excess galaxies entering (or exiting) the source sample due to magnification do contribute to the lensing shear, the 
		measured boost factor is biased by
		\begin{align}
			\frac{{\delta b_F}}{b_F}(r_p)=\frac{1}{b_FN}\int dz_lp(z_l)f_k(\chi_l)^{-2}\int dz_p p(z_p)\Sigma_{c}^{-2}(z_l,z_p)\int dz_sp(z_s|z_p)
				\left[\mean{\delta_l\kappa'_s+\kappa'_l\kappa'_s}(r_p)\right]
		\end{align}
		$\alpha_s\approx0.55$ for our source sample. The $\delta_l\kappa'_s$ term can be approximated as  $2(\alpha_s-1)\frac{\Sigma_{gm}}{b_F 
		\Sigma_\text{crit}}$ in terms of lens $\Sigma_{gm}$. For the lens-source pairs in our
      analysis, the effective $\Sigma_\text{crit}\sim4600[h\Msun/pc^2]$ 
      	and 
		thus the contamination from 
		the $\delta_l\kappa_s'$ term is order $10^{-4}\times\Sigma_{gm}$. $\kappa'_l\kappa'_s$ is similar to the term derived 
		earlier in the section, 
		 except for the $\Sigma_\text{crit}$ factor and will be order $10^{-5}$--$10^{-6}$.

	\subsection{Bias in $E_G$}
        The bias in $E_G$ can be estimated as (here we ignore the bias in $f$ as the bias in multipole moments is order 0.1\% or lower compared to $
        \sim5\%$ or higher uncertainty in $f$)
		\begin{align}
			\frac{\Delta E_G}{E_G}(r_p)=&\frac{\widehat{b}_F}{b_F}\frac{\widehat{\Delta\Sigma}}{\Delta\Sigma}
									\frac{w_{gg}}{\widehat{w}_{gg}} (r_p)-1
								\approx \frac{\langle\Sigma_\text{crit}\gamma_t\kappa'\rangle}{\Delta\Sigma}+\frac{\mean{\delta_l\kappa'_s}}{b_F}
								+\frac{\mean{\kappa'_l\kappa'_s}}{b_F}
								+\frac{w_{g\kappa'}}{w_{gg}}-\frac{w_{\kappa'\kappa'}}{w_{gg}}
								\approx 10^{-2}
		\end{align}	
		Here for brevity, we omitted showing various integrals involved in the terms.
		Since the boost factor is multiplied to the lensing signal, it also affects \eg\ as a multiplicative factor. 
		The final corrections applied to the $E_G$ measurement are shown in Fig.~\ref{fig:EG_corrections} ($C_\text{lens}$), 
		for the case of LOWZ sample. 
		
\section{Testing the $E_G$ corrections using different mock catalogues}\label{append:hod_test}
	In this section we test the theoretical uncertainties associated with corrections computed from mocks, especially $C_{nl}$, 
	using mock catalogues with different HOD models. 
    For this purpose,
	we construct light cones from the BigMultiDark Planck simulation \citep[BigMDPL;][]{Klypin2016} using all the
	available snapshots in the range $0.16<z<0.36$. This is a simulation of a flat $\Lambda$CDM
    model, with Planck 2013 
	cosmological parameters \citep{planck2014}, $\Omega_m=0.307$, $\Omega_\Lambda=0.693$, $h=0.678$ and
	$\sigma_8=0.829$. The BigMDPL simulation has 3840$^3$ particles with a mass resolution of 2.4$\times10^{10}$ $h^{-1}M_\odot$ 
	and a box size of 2.5 $h^{-1}$Mpc. These features allow the creation of light cones with a
    volume comparable to BOSS with resolved 
	dark matter halos in the mass range predicted for the LOWZ and CMASS galaxies. Dark matter halos are 
	defined using the Robust Over-density Calculation using K-Space Topologically Adaptive Refinement halo finder 
        \citep[\textsc{rockstar};][]{Behroozi2013}.

        We use the above light cones to produce galaxy mocks that reproduce the clustering at different redshifts 
	and the radial selection function of the observed sample. 
    In order to populate the dark matter halos of the simulation we implement the subhalo abundance
	matching (SHAM) used by \citet{Torres2016}. The scatter is included using the maximum circular velocity 
	over the whole history of the (sub)halo, $V_\text{peak}$, using the relation
        \begin{equation}
          \label{eq:ham}
          V_\text{peak}^{scat}=(1+\mathcal{G}(0,\sigma_{\textsc{sham}}))V_\text{peak},
        \end{equation}
        where $\mathcal{G}$ is a random gaussian number from a distribution with mean 0 and standard deviation 
	$\sigma_{\textsc{ham}}$. We sort all (sub)halos using $V_\text{peak}^{scat}$ and select objects from the largest
	velocity and continuing down until we reach the number density of the observed sample. The scatter between halos 
	and galaxies is fixed using the projected correlation function of the LOWZ data between $2<r_p<30$ $h^{-1}$ Mpc for 
	3 different redshift bins in the range $0.16<z<0.36$.  

        We do not distinguish between host halos and subhalos in the first light-cone. This produces a LOWZ mock with 12.5\% of galaxies living in subhalos. In order to produce a different HOD, we include an additional parameter which 
	models the fraction of galaxies living in subhalos, $f_{sat}$. It will force the model to populate a desired
      fraction of subhalos with galaxies, rather than the value that naturally arises through the
      procedure described above. 
	After including the new parameter, we produce 4 different light-cones, first increasing the fraction of satellites by 10 
	and 15 percent and then decreasing by the same factors. In all cases, we fix the scatter to reproduce the 
	projected correlation function of the LOWZ sample.
	
\begin{figure*}
        	\centering
		\begin{subfigure}{.45\columnwidth}		
         	\includegraphics[width=\columnwidth]{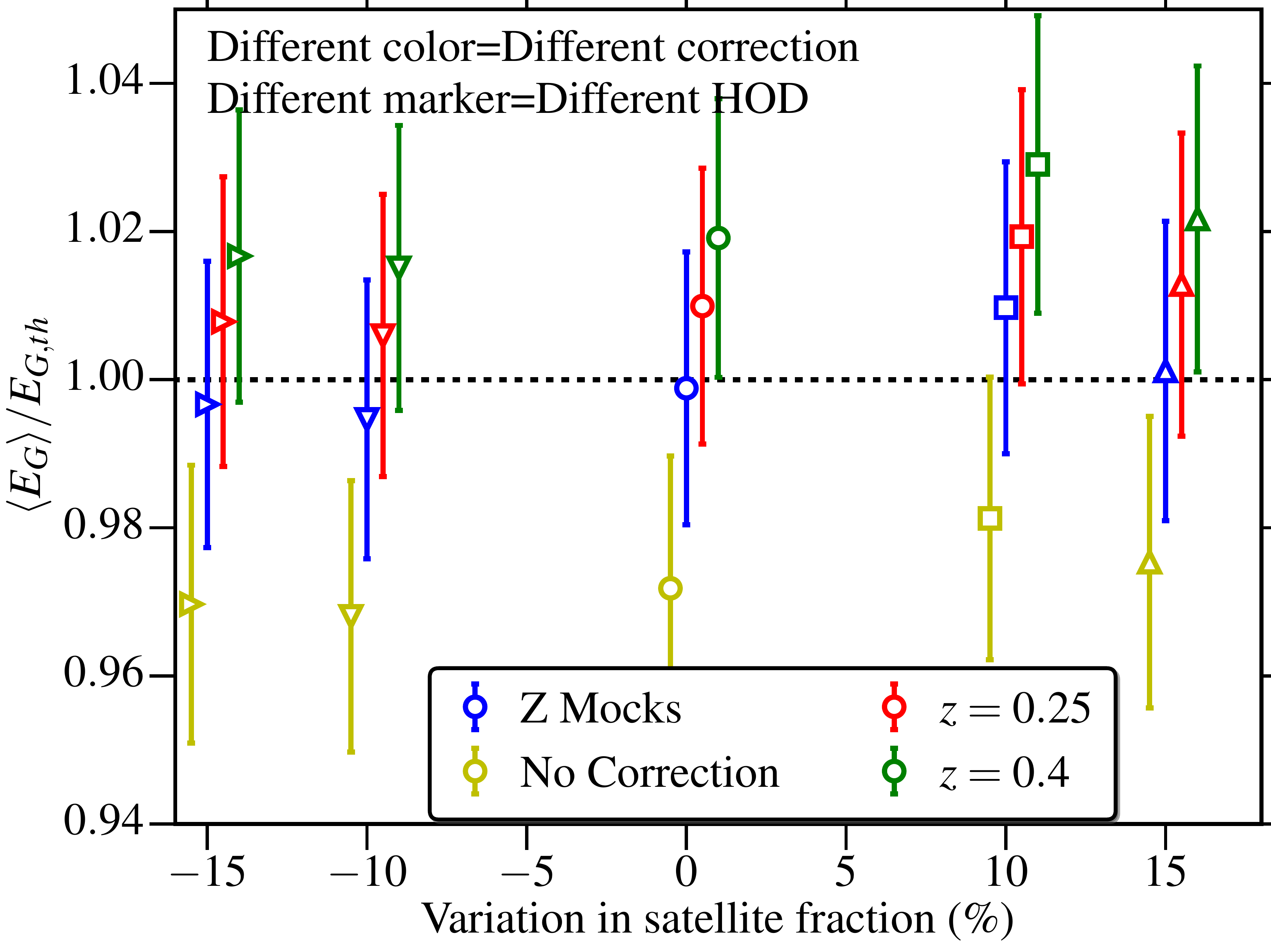}
	         \caption{
				}
    	     \label{fig:EG_correction_mocks_hod}
	     \end{subfigure}
	     \begin{subfigure}{.45\columnwidth}		
         	\includegraphics[width=\columnwidth]{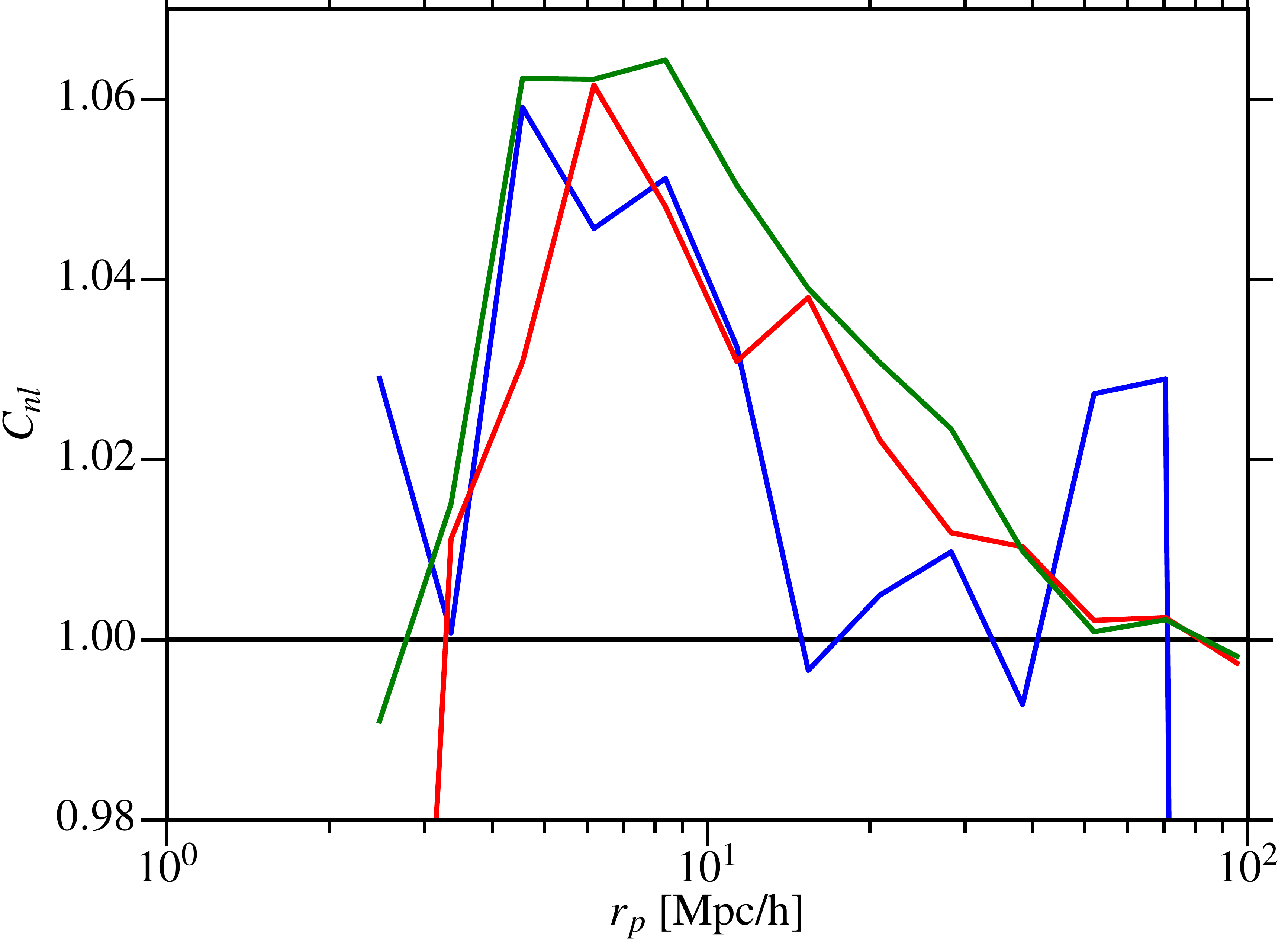}
	         \caption{
				}
    	     \label{fig:EG_correction_comparison}
	     \end{subfigure}
	    \label{fig:EG_correction_mocks}
	     \caption{a) Ratio of theoretical prediction of \eg\, $E_{G,th}$ with $E_G$ measured using mocks with different HODs 
	     (variable satellite fraction) and different sets of 
	     corrections applied. `Z mocks' correction is computed from MDPL mocks with default satellite fraction ($0\%$ variation).
	     b) $C_{nl}$ correction computed from different mocks that are used for corrections in a) (same color scheme in both panels). The 
	     differences
	     in the $C_{nl}$ result in variations \mean{\eg} at order of $1-2\%$ and thus we add $2\%$ systematic uncertainty resulting from these 
	     corrections.
	     }
        \end{figure*}

	In addition to $C_{nl}$ measured using simulations presented in section~\ref{ssec:data_simulations}, 
	we also compute $C_{nl}$ from one of BigMDPL mocks (labeled as `Z Mocks'). 
	We then apply the three sets of $C_{nl}$ to mocks with different 
	HODs to check if we recover $E_G$ accurately. In Fig.~\ref{fig:EG_correction_mocks_hod}, we show the relative bias 
	in the corrected $\mean{E_G}$ relative to the theoretical prediction. Applying no corrections, $\mean{E_G}$ is 
	biased low by $\sim3\%$. Applying the correction from same MDPL mocks, we recover the
    $\mean{E_G}$ to within 
	$\lesssim0.5\%$ independent of the HOD. However, applying the correction from different
    simulations leads to an overcorrection, and 
	$\mean{E_G}$ is biased high. As we show in Fig.~\ref{fig:EG_correction_comparison}, this is primarily driven by
	differences in the non-linear physics, which leads to different $C_{nl}$, across different
    simulations. The corrections used 
	for our main results, from the $z=0.25$ snapshot with the fiducial HOD (no variation in
    satellite fraction), recovers $\mean{\eg}$ to within $\sim2\%$ for all HODs.  Thus we add $2\%$ systematic uncertainty due to 
    these corrections to our results.

	\section{Photo-z Lensing bias estimates using clustering redshifts}\label{append:clustering_pz}

		In this section, we present the estimates of the true redshift distribution $p(z_s)$ for our photometric 
		source sample, closely following the 
		formalism developed by \citet{Menard2013,van_Daalen2017}. 
		We use a random subsample of our source sample containing $\sim2\times10^6$ galaxies and split it into
		40 photo-$z$ bins containing approximately equal numbers of galaxies. These samples are then cross-correlated 
		with the BOSS spectroscopic sample (LOWZ+CMASS), which is split into bins of size $\delta
        z=0.05$ in the range $z\in[0.1,0.8]$ and two additional bins in the ranges $z\in[0,0.1]$ and
        $z\in[0.8,1]$. The BOSS 
        sample that we use does not have many galaxies for $z\in[0.8,1]$, where the constraints will be very noisy. We tested our
        results by omitting this bin and the qualitative trends in the lensing bias presented later in this section do not 
        change, though omitting that bin does change the actual values of the bias in the lensing signal by
        $\sim 4\%$. This is because the galaxies at these redshifts have low $\Sigma_\text{crit}$ and hence have larger weight in the 
        lensing measurement, and $\Sigma_\text{crit}$ at larger source redshifts is 
        relatively less affected by small changes in redshifts.
		
		To compute auto- and cross-correlations between the spectroscopic samples, we use the standard Landy-Szalay 
		estimator \citep{Landy1993}, but for cross-correlations involving the source sample we use the sub-optimal
        estimator as we do not
		have a good estimate for the on-sky selection function of the source catalog and hence we
        lack random catalogs for that sample:
		\begin{align}
			w_{s_i-s_j}(\theta)=&\frac{\sum_{\theta}(D_{s_i}-R_{s_i})(D_{s_j}-R_{s_j})}{\sum_{\theta}R_{s_i}R_{s_j}}
			=\frac{\sum_{\theta}(D_{s_i}D_{s_j}-R_{s_i}D_{s_j}-D_{s_i}R_{s_j}+R_{s_i}R_{s_j})}
				{\sum_{\theta}R_{s_i}R_{s_j}}\\
			w_{p_i-s_j}(\theta)=&\frac{\sum_{\theta}D_{p_i}(D_{s_j}-R_{s_j})}{\sum_{\theta}D_{p_i}R_{s_j}}
				=\frac{\sum_{\theta}(D_{p_i}D_{s_j}-D_{p_i}R_{s_j})}{\sum_{\theta}D_{p_i}R_{s_j}}
		\end{align} 
		Here the subscript ${p_i}$ refers to the $i^{th}$ photometric sample, $s_i$ refers to the $i^{th}$
        spectroscopic sample and $\sum_\theta$ indicates that summation is carried over pairs with
        separation $\theta$ within 
        the limits of the given bin. 
        Then we integrate the correlation functions to obtain
		\begin{align}
			W_{a-b}=\int_{\theta_\text{min}}^{\theta_\text{max}} w_{a-b}(\theta)\mathcal W(\theta) \theta d\theta
		\end{align}
		We use the weight function $\mathcal W(\theta)=1/\theta$, as suggested by \cite{Menard2013}. 
		We use $\theta_\text{min}=0.03$ degree and $\theta_\text{max}=0.5$ degree. These choices are primarily dictated 
		by the 
		noise and systematic bias in the measured cross-correlation functions, which are largely determined by
        the SDSS number density
		at small $\theta$ and our use of the sub-optimal estimator at large $\theta$, which
        increases the variance \citep[][there is also some evidence of 
		biases due to selection function effects in some bins]{Singh2016cov}.
		
		The cross correlation between a photometric sample, $p$, and spectroscopic sample $s_i$ can be modeled as
	\begin{equation}\label{eq:W_cross_predict}
		W_{p-s_i}=\sum_{j=1}^{N_s}b_{R,p-s_j} f_{p,s_j} W_{s_i-s_j}.
	\end{equation}
	$N_s$ is the number of spectroscopic subsamples, $f_{p,s_j}\approx p(z_s)\Delta z_s$ is the fraction of 
	the photometric sample 
	within the redshift range of spectroscopic bin $j$ and the bias ratio $b_{R,p-s_j}$ is defined as
	\[b_{R,p-s_j}=
	\begin{cases}
		\frac{b_{p,s_j}}{b_{s_j}}	& z_{s_j}\le z_{s_i} \\
		 1 & \text{otherwise}.
	\end{cases}\]
	$b_{p,s_j}$ is the (integrated) bias of photometric galaxies that lie in redshift bin $z_{s_j}$ and 
	$b_{s_j}$ is the bias of spectroscopic galaxies (measured by fitting the linear$+$halofit matter correlation 
	function to 
	$w_{s_j-s_j}$) in the same bin.
	If $z_{s_j}\le z_{s_i}$, galaxies at $z_{s_j}$ are either correlated with galaxies at $z_{s_i}$ 
	($z_{s_j}=z_{s_i}$) or they are at lower redshift and are lensing the galaxies at higher redshift $z_{s,i}$. When
	$z_{s_j}>z_{s_i}$, galaxies at $z_{s_i}$ act as lens and hence the correlations depend only on
    the bias of the spectroscopic galaxies in $z_{s_i}$ and the ratio is 1 (this is strictly not true as lensing effects 
    (convergence) also depend on the luminosity function of the source sample as shown in appendix~\ref{append:lensing_EG}, 
    but we ignore that effect here).
    We also stress here that bias going into our estimation is some effective bias integrated over the scales and can also 
    contain effects from non-linear cross correlation coefficients between spectroscopic and photometric samples (similar to 
    $r_{cc}$). 
		
	In our fitting 
	procedure, we leave $b_{p,s_j}$ as a free parameter,  though we do put hard lower and upper limits on the values it 
	can take and also assume that it follows some simple redshift evolution. For hard limits, we use 
	$b_\text{min}<b_{p,s_j}<3\,\forall\, p,s_j$, where the upper limit is fixed to 3 and for the lower limit we try 3 different 
	values $b_\text{min}=[0 , 0.4, 0.7]$. $b_\text{min}\approx0.7$ is motivated by the lower limit of halo bias in 
	simulations \citep[e.g.,][]{2010ApJ...724..878T}. 
    However, 
	to accommodate any effects of non-linear physics,  
	we also do the analysis with different lower limits of galaxy bias.
	To fit for the redshift evolution, we try three different functions: constant, linear and power-law, with 
	two free parameters, $b_0$ and $b_1$, to be fitted,
	\begin{align}\label{eq:bias_evolution}
		b_p(z)=&b_0\left(\frac{1+z}{1+\overline{z}_p}\right)^{b_1}\\
		b_p(z)=&b_0\left[{1+b_1(z-\overline{z}_p})\right]
	\end{align}
	where $b_1=0$ for constant bias and $\overline{z}_p$ is the mean photometric redshift of the
    given photometric subsample. 
	
	For each photometric redshift bin, we adopt a Gaussian likelihood as
	\begin{align}
		L_W=\frac{1}{\sqrt{|\Sigma|}}\exp\left(-\frac{1}{2}(\widehat{W}_{p-s_i}-W_{p-s_i})^T\Sigma^{-1}(\widehat{W}_{p-
		s_i}-W_{p-s_i})\right)
	\end{align}
	We fit for $f_{p,s_j},b_0,$ and $b_1$. 
	To fit for the fractional distribution of galaxies, $f_{p,s_j}$, we use two separate methods. 
	In the first method, we assume the $p(z_s)$ is Gaussian and integrate the 
	$p(z_s)$ within a bin to obtain $f_p$. In this case, the fitting parameters are the mean and variance of the 
	Gaussian $p(z_s)$.
	However, using a Gaussian function for the $p(z_s)$ is a simplification that may not be justified
    in reality. So we also use a non-parametric fitting procedure where each of the $f_{p,s_j}$ are 
	free parameters.
	In this case, we require that $-f_l<f_{p,s_j}<1+f_l$. We test the values $f_l=[0,0.05,0.1]$, where
	values $>0$ are necessary to accommodate shot noise.
	While the second method is in principle the most general form for $f_p$, the
	results have a higher variance and can be more susceptible to the effects of bias-$f_p$ degeneracy.
	When computing the photo-z bias in lensing, any bins $j$ with best-fitting $f_{p,s_j}<0$ are set to zero (since bins 
	with no galaxies due to 
	shot noise will not contribute anything to lensing).

	In addition, we also require that the sum of 
	$f_{p,s_j}$ ($\sum_j f_{p,s_j}$) should be close to 1. For this we add an additional term in the
    likelihood as 
	\begin{align}
		\mathcal L=&L_W\frac{1}{\sqrt{N_{p,f}}}\exp\left(\frac{-1}{2}\frac{N_{p,f}^2
			(\sum_j f_{p,s_j}-1)^2}{N_{p,f}}\right)\\
			N_{p,f}=&\frac{500}{\overline z_p}
	\end{align}
	Where $N_{p,f}$ should ideally be the number of galaxies in the given photometric sample. However,
	since the redshift extent of the spectroscopic sample is less than that of the photometric sample, we loosen 
	the likelihood for samples at higher photometric redshift, hence the dependence on $z_p$. 
	500 was chosen so that $500/\text{min}(z_p)$ is of order 50,000, which is the number of galaxies in each photometric bin.
	
	Once we have the spectroscopic redshift distribution for given photometric samples, we can
    estimate the bias, $B_L$ 
	in measured $\widehat{\Delta\Sigma}$ as \citep{Nakajima2012} (note that we use different notation than \cite{Nakajima2012} as $b_z$ can be 
	confused with redshift-dependent bias)
	\begin{equation}
		B_L=\frac{\widehat{\Delta\Sigma}}{\Delta\Sigma}=\frac{\int dz_l\, p(z_l)D(z_l)^2 b(z_l) f_k(\chi_l)^{-2}\int dz_p\, p(z_p)w_{ls}(z_l,z_p)
		\int dz_s \, p(z_s|z_p)\frac{\Sigma_\text{crit}(z_l,z_p)}{\Sigma_\text{crit}(z_l,z_s)}}
		{\int dz_l\, p(z_l)D(z_l)^2 b(z_l) f_k(\chi_l)^{-2}\int dz_p\, p(z_p)w_{ls}(z_l,z_p)
		\int dz_s\,p(z_s|z_p)}
	\end{equation}
	$z_l$ is the lens redshift, $z_p$ is the photometric redshift, $z_s$ is the true source redshift and 
	$\int dz_s \, p(z_s|z_p)=f_{p,s}$. $D(z_l)$ is the 
	growth function at lens redshift, $b(z_l)$ is the galaxy bias, $f_k(\chi_l)$ is the comoving transverse distance at lens redshift (accounts for the fact that effective 
	number of source galaxies entering a fixed comoving distance bin decreases at higher lens
    redshift -- \citealt{Nakajima2012})
   and  $w_{ls}$ is the weight as described in Eq.~\eqref{eq:delta_sigma_wt}. For our computation,
   we assume that the clustering of lens galaxies is redshift independent and hence $b(z_l)\propto
   D(z_l)^{-1}$ and $\Delta\Sigma(z_l) \propto D(z_l)$.

\begin{figure*}
	\includegraphics[width=.8\columnwidth]{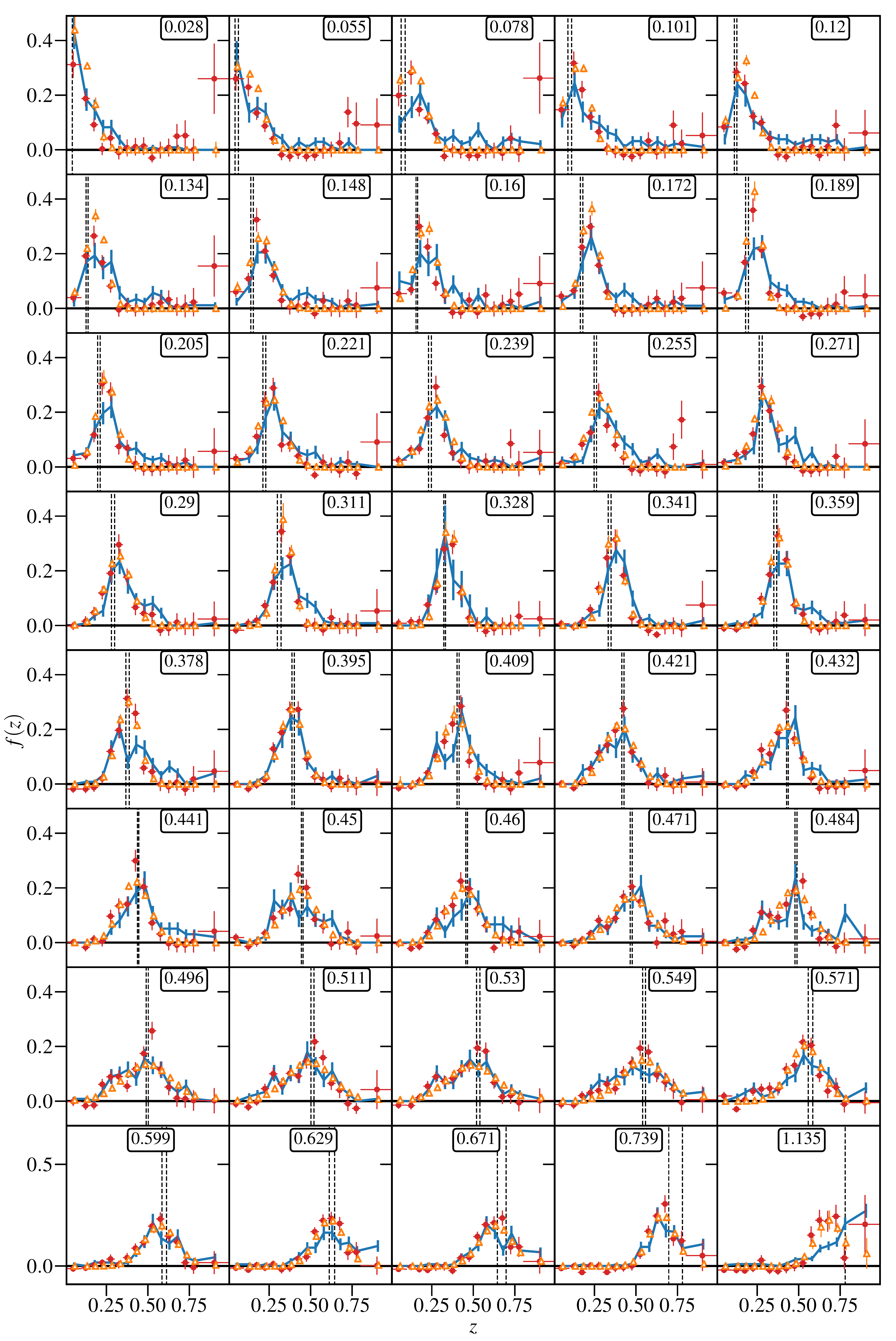}
		\caption{Redshift distribution estimates $f(z)$ for different photometric redshift bins,
          where each panel represents a given photo-z selected sample. 
			Red and orange points are for the fits with $f(z)$ in each bin as a 
			free parameter and assuming a Gaussian redshift distribution, respectively.  For both red and orange 
			curves in this plot we use $b_\text{min}=0.7$; for the red points, $f_l=0.05$. The blue curves
            show the $f(z)$ obtained
			using spectroscopic redshifts from the representative calibration sample.  
			}
        \label{fig:fz_cl}
    \end{figure*}

	In Fig.~\ref{fig:fz_cl}, we show the redshift distribution (fraction of galaxies in each bin) 
      estimated assuming a Gaussian model for $p(z_s)$ (blue 
	points) and when using the non-parametric $f_p$ model (red points). We also show the $f_p$ obtained using a
    representative calibration sample
	with spectroscopic redshifts from \cite{Nakajima2012}. $f_p$ obtained using the Gaussian model is less noisy, but for any 
	given photometric
	sample, the bins in true redshift are strongly correlated. In the case where the $f_p$ are treated as independent 
	parameters, the 
	correlations across bins are not significant and the distributions appear to capture more features, though to some 
	extent this can be caused by noise in clustering measurements and the degeneracies with the galaxy bias.
	
\begin{figure*}
    	\centering
	\begin{subfigure}{.70\columnwidth}		
    	\includegraphics[width=1\columnwidth]{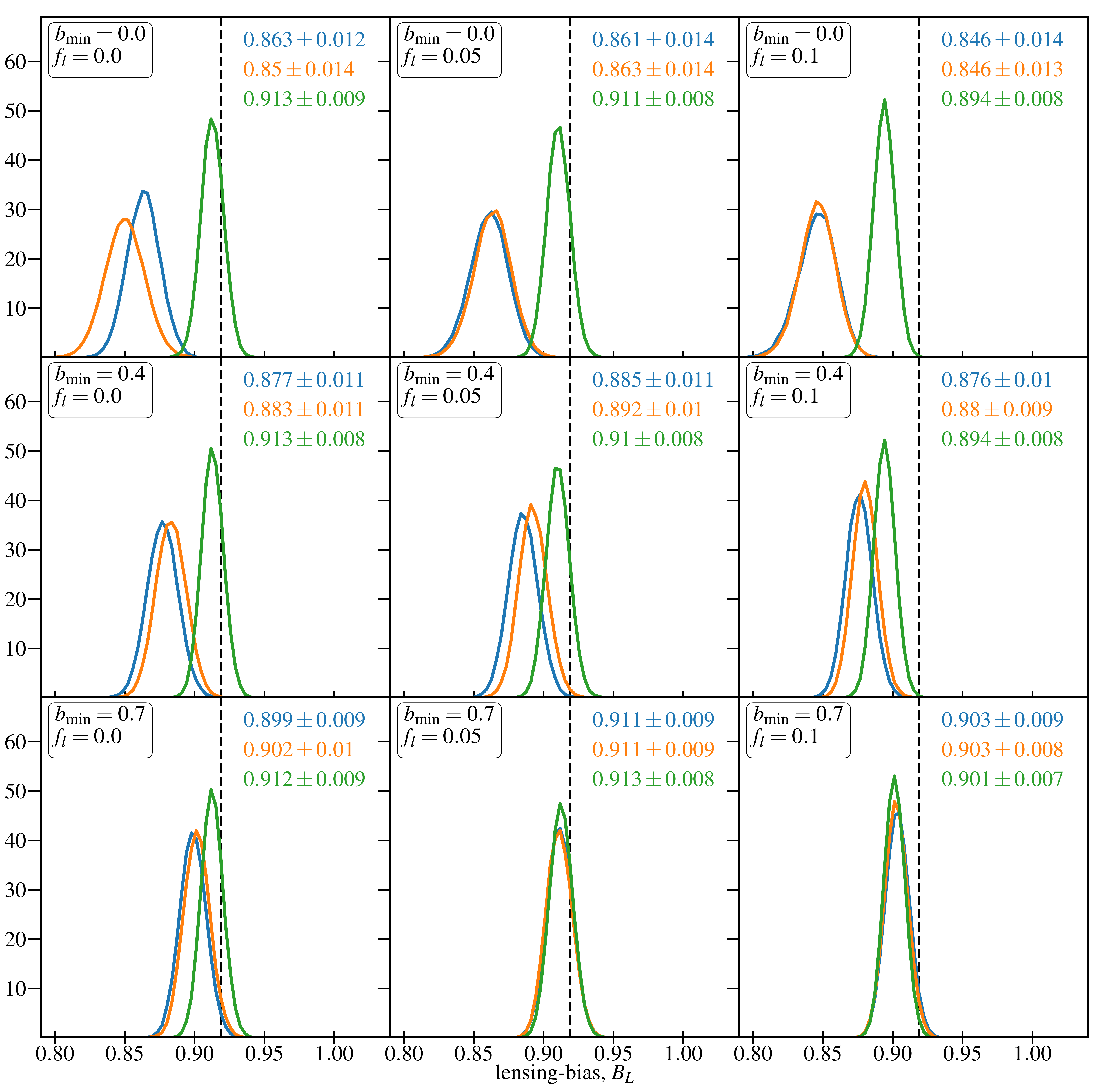}
		\label{}
		\caption{Non-Parametric $p(z)$}
	\end{subfigure}
	\begin{subfigure}{.255\columnwidth}		
    	\includegraphics[width=1\columnwidth]{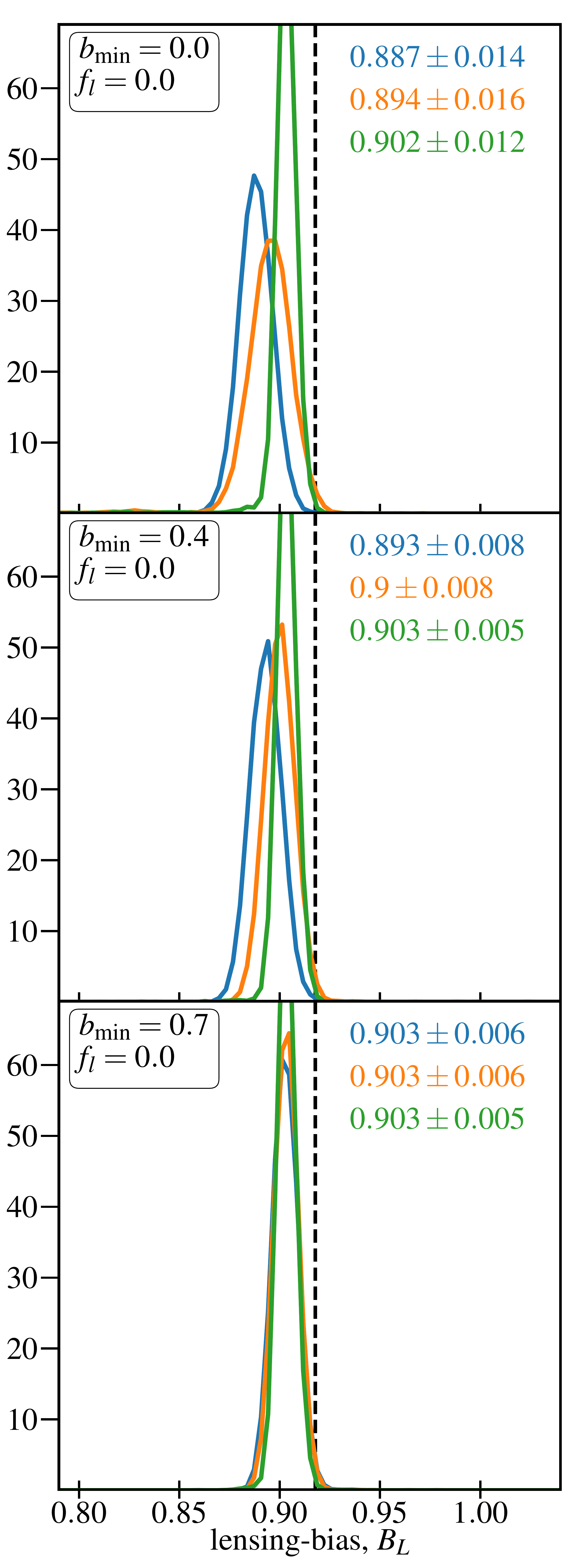}
		\label{}
		\caption{Gaussian $p(z)$}
	\end{subfigure}
		\caption{a) Probability distribution of bias in the lensing signal, $B_L$, due to photometric redshifts, 
			for different choices of lower limit on bias and redshift fraction.  
			The vertical dashed black lines show the estimate using spectroscopic
			redshifts in the representative calibration sample.
			The blue lines/points are assuming power law model for the bias-redshift relation, orange color is for linear model 
			while green is for constant bias with redshift.
			When bias is allowed to evolve with redshift, there is strong dependence of the inferred bias on the lensing signal 
			on the priors chosen, though the choice of model for the bias-redshift relation does not affect
            it too strongly. In all cases, the difference between the lensing bias from the clustering
            redshift method and that derived using spectroscopic redshifts ($\sim0.92$) is
            $\lesssim10\%$.
            b) Same as a), using gaussian model for $p(z)$. Green curves show the estimate when we assume 
            galaxy bias to be constant with redshift ($b_1=0$ in Eq.~\eqref{eq:bias_evolution}).
}

        \label{fig:pz_bias}
    \end{figure*}

	In Fig.~\ref{fig:pz_bias} we show the bias in the galaxy-galaxy lensing measurement due to photometric redshift
    bias and scatter, calculated using
	the redshift distributions estimated from the clustering analysis. 
    We show the estimates using different bias 
	and $f_p$ models along with different priors on the galaxy bias and $f_p$. Our results are not very sensitive to the 
	bias evolution models, likely due to the relatively narrow redshift bins. 
	However, in the case where the $f_p$ values are treated as free parameters, the results are sensitive to the priors 
	on $f_p$ 
	and the bias, especially the lower limit on bias. This is primarily caused by the degeneracy
    between the bias and $f_p$, along with the noise in the correlation function measurements that makes it difficult
    to infer so many free parameters. We attempt to overcome this difficulty using the Gaussian $p(z_s)$. The 
    results in this case are less sensitive to priors on galaxy bias. However, the $S/N$ in the
    estimate of the lensing bias 
    is not improved by much, which suggests that we do not gain much more information by constraining
    $p(z_s)$ (the $f(z_s)$ 
    obtained in this case are correlated across bins). As a result we do not attempt to use more complicated functions 
    than gaussian which may capture more features of $p(z_s)$.  

	For the physically-motivated choice of priors, $b_\text{min}\in[0.4,0.7]$ and $f_l\in[0.05,0.1]$,
    our results are consistent
	(within $\lesssim 5\%$) with the bias in the lensing signal estimated using the representative
    spectroscopic calibration
    sample.  This value has been 
	used in this and previous works to correct the lensing measurements for the impact of photo-$z$
    bias and scatter.
	
	Our results suggest that more work is required to carefully choose the priors 
	and functional forms for $f_p$ to further develop the clustering redshift method (as used in this work) 
	for calibration of the galaxy-galaxy lensing signal. However, our results demonstrate that the bias in our galaxy-galaxy 
	lensing signals due to photometric redshift bias and scatter 
	is well constrained at the $\lesssim5\%$ level and hence is not likely the full explanation for $\sim15\%$ discrepancy
	observed between the \eg\ measurements and the prediction from the Planck \lcdm\ model.
	
\label{lastpage}
\end{document}

%% file: def.tex
\newcommand{\wgg}{\ensuremath{w_{gg}}}
\newcommand{\ugg}{\ensuremath{\Upsilon_{gg}}}
\newcommand{\ugm}{\ensuremath{\Upsilon_{gm}}}
\newcommand{\dsigma}{\ensuremath{\Delta \Sigma}}
\newcommand{\eg}{\ensuremath{E_G}}
\newcommand{\mean}[1]{\ensuremath{\left\langle #1 \right\rangle}}
\newcommand{\rcc}{\ensuremath{r_\text{cc}}}
\newcommand{\lcdm}{\ensuremath{\Lambda\text{CDM}}}
\newcommand{\wgm}{\ensuremath{w_{gm}}}

\newcommand{\xigg}{\ensuremath{\xi_{gg}}}

\newcommand{\mpch}{\ensuremath{h^{-1}\text{Mpc}}}

\def\reff@jnl#1{{\rm#1\/}}

\def\aj{\reff@jnl{AJ}}                  
\def\araa{\reff@jnl{ARA\&A}}            
\def\apj{\reff@jnl{ApJ}}                        
\def\apjl{\reff@jnl{ApJ}}               
\def\apjs{\reff@jnl{ApJS}}              
\def\apss{\reff@jnl{Ap\&SS}}            
\def\aap{\reff@jnl{A\&A}}               
\def\aapr{\reff@jnl{A\&A~Rev.}}         
\def\aaps{\reff@jnl{A\&AS}}             
\def\baas{\reff@jnl{BAAS}}              
\def\jrasc{\reff@jnl{JRASC}}            
\def\memras{\reff@jnl{MmRAS}}           
\def\mnras{\reff@jnl{MNRAS}}            
\def\physrep{\reff@jnl{Phys.Rep.}}
\def\pra{\reff@jnl{Phys.Rev.A}}         
\def\prb{\reff@jnl{Phys.Rev.B}}         
\def\prc{\reff@jnl{Phys.Rev.C}}         
\def\prd{\reff@jnl{Phys.Rev.D}}         
\def\prl{\reff@jnl{Phys.Rev.Lett}}      
\def\pasp{\reff@jnl{PASP}}              
\def\pasj{\reff@jnl{PASJ}}              
\def\skytel{\reff@jnl{S\&T}}            
\def\solphys{\reff@jnl{Solar~Phys.}}    
\def\sovast{\reff@jnl{Soviet~Ast.}}     
\def\ssr{\reff@jnl{Space~Sci.Rev.}}     
\def\nat{\reff@jnl{Nature}}             

\newcommand{\Msun}{M_{\odot}}

\newcommand{\beq}{\begin{equation}}
\newcommand{\eeq}{\end{equation}}
\newcommand{\beqa}{\begin{eqnarray}}
\newcommand{\eeqa}{\end{eqnarray}}



%% file: GrowthM.tex
We performed the RSD analysis as described in Section~\ref{ssec:formalism_rsd}. For every sample,
monopole and quadruple moments in each jackknife region are fit independently using our perturbation theory
model.
The measurements and model fits are shown in figure~\ref{fig:rsd_multipole} and the growth rate measurements are presented in 
table~\ref{tab:params}, with the results being 
 consistent with $\Lambda$CDM predictions to within
\referee{$\sim 1\sigma$}. 

Our RSD analysis is  
 performed with a fixed cosmology because the constraints on cosmological parameters from 
Planck are tight enough that using Planck priors is nearly equivalent to fixing the cosmology. 
Our measurements of the growth rate are consistent with previous measurements using the SDSS sample \citep{AlamRSD2015,
2014MNRAS.440.2692S, 2014MNRAS.439.3504S, 2016arXiv160703148S, 2016arXiv160703150B,
2016arXiv160703143G, 2016arXiv160703147S,  Boss2016combined}.
We do not marginalize over the
Alcock-Paczynski parameter and hence obtain smaller error than when doing the full shape RSD analysis (see for
example \citealt{AlamRSD2015}). 
\cite{2014MNRAS.440.2692S} reported $f\sigma_8(z=0.32)=0.48\pm0.10$
and $f\sigma_8(z=0.57)=0.42\pm0.045$ using the SDSS DR11 sample, which is consistent with our LOWZ
($f\sigma_8(z=0.27)= 0.44$) and CMASS ($f\sigma_8(z=0.57)=0.42$). 
A more detailed comparison of RSD modeling used in this paper with several other methods is shown in \cite{ Boss2016combined}. 
\cite{2016MNRAS.460.4188G} presented one of the first RSD analysis
 which uses both power spectrum and bi-spectrum and their reported
 $f\sigma_8$ is consistent with ours for both LOWZ and CMASS sample. Our LOWZ measurement of $f\sigma_8$ is 0.5$\sigma$ lower than \cite{2016MNRAS.460.4188G}
 with fixed Alcock-Paczynski parameter but higher than what they measured after marginalizing over
 Alcock-Paczynski parameters. The lower $f\sigma_8$  after AP marginalization is dominated by the
 information in the position of the BAO peak. We do not use the BAO scale in our RSD analysis
and hence do not allow extra freedom on top of the $\Lambda$CDM background when estimating
 $f\sigma_8$. We also note that in our model the growth rate ($f$) and additional velocity dispersion ($\sigma_{FOG}$) are uncorrelated with each other for small values of $\sigma_{FOG}$ but positively correlated for large values of $\sigma_{FOG}$.  Therefore galaxy sample with stronger finger-of-god effect will show stronger effect on the growth rate constraint when marginalized over $\sigma_{FOG}$. We found that our $f\sigma_8$ constraint for LOWZ sample are not strongly affected by marginalizing over $\sigma_{FOG}$ but for the CMASS sample the error on growth rate increases by a factor of 2 when marginalizing over $\sigma_{FOG}$ compared to when it is fixed (the mean value of $f$ shifts by less than $0.2\sigma$). This can plausibly be explained by the fact that the volume occupied by CMASS sample is bigger and therefore gives a stronger constraint on $f\sigma_8$ which makes it easier to see the impact of marginalizing over $\sigma_{FOG}$. 

\referee{The systematic uncertainty in our RSD model is estimated to be $\sim2\%$ as reported in \cite{Alam2016} while using scales above 25 \mpch. But, our sample extends to slightly lower redshift compared to previous analysis and hence we decided to use a slightly higher minimum scale in our RSD fits hoping to keep the systematic error at the same level. We used scales above 35 \mpch\,  in our measurement of $\beta$} 
whereas our final $E_G$ measurements extends to scales  
below 35 \mpch. 
Ideally one would want to perform measurement of $\beta$ using the same scales. 
Unfortunately the modeling technique used in the current measurement of $\beta$ is not accurate
enough to extend it to smaller scales. The $\beta$ consists of two quantities growth rate and
bias. We do account for the fact that bias will be scale dependent and different at smaller scales
compared to the large scale measurement through a correction factor $C_{nl}$ (see Section~\ref{ssec:formalism_corrections} for details). But we have an inherent assumption that the growth rate measured using larger scales is constant and applicable for smaller scales. This makes the current measurements of $E_G$ slightly weaker than its full potential which should be improved upon in the future measurements with better RSD modeling to smaller scales.